\documentclass[11pt,a4paper]{article}
\usepackage{epsfig,graphicx,amsfonts,amsmath}

\usepackage{cite}
\RequirePackage{mathrsfs}
\usepackage{color}

\newlength{\dinwidth}
\newlength{\dinmargin}
\def\fig#1{{Fig.~(\ref{#1})}}
\def\eq#1{{Eq.~(\ref{#1})}}
\newcommand{\Le}{\left(}
\newcommand{\Ra}{\right)}

\newcommand{\beq}{\begin{equation}}
\newcommand{\eeq}{\end{equation}}
\newcommand{\beqar}{\begin{eqnarray}}
\newcommand{\eeqar}{\end{eqnarray}}
\newcommand{\D}{\partial}

\newcommand{\ph}{\varphi}

\newcommand{\om}{\omega}

\newcommand{\tu}{\textsl{u}}
\newcommand{\tv}{\textsl{v}}

\newcommand{\rom}[1]{\uppercase\expandafter{\romannumeral #1\relax}}

\date{}
\begin{document}

\title {{~}\\
{\Large \bf CPTM symmetry, closed time paths and cosmological constant problem in the formalism of extended manifold}}
\author{ 
{~}\\
{\large 
S.~Bondarenko$^{(1) }$
}\\[7mm]
{\it\normalsize  $^{(1) }$ Physics Department, Ariel University, Ariel 40700, Israel}\\
}

\maketitle
\thispagestyle{empty}

\begin{abstract}
 The problem of the cosmological constant is considered in the formalism of an extended space-time consisting of the extended classical solution of Einstein equations. 
The different regions of the extended manifold are proposed to be
related by the charge, parity, time and mass (CPTM) reversal symmetry applied with respect to the metric fields of the manifolds. 
There are interactions between the points of the extended manifold provided by scalar fields present separately in the different 
patches of the extended solution. 
The value of the constant is obtained equal to zero at the classical level due the mutual contribution of the fields in the vacuum energy,
it's non-zero value is due the quantum interactions between the fields.
There are few possible scenario for the actions of the fields
are discussed. Each from the obtained variants is similar to the closed time path approach of non-equilibrium condensed matter physics and
among these possibilities for the
closed paths, there is a variant of the action equivalent to the formalism of Keldysh. 
Accordingly, we consider and shortly discuss the application of the proposed formalism to the problem of smallness of the cosmological constant and singularities problem.  

\end{abstract}

\section{Introduction}
\label{S1}

  Following to  \cite{Ser1,Ser2}, we investigate the appearance of the cosmological constant in the formalism of an extended classical solution of Einstein equations 
as a consequence of the quantum interaction between the parts of the extended manifold. The metrics of the separated manifolds
of the extended solution  are related by the discrete reversal charge, parity, time~and mass  
(CPTM) symmetry which preserves the form of the metric $g$ at the case of the zero cosmological constant.
The easiest way to clarify this construction is to consider as an example the light cone coordinated $u,v$
or corresponding Kruskal--Szekeres coordinates \cite{Kruskal} in the Schwarzschild's spacetime defined for the whole space-time solution.
In this case,  the extended CPTM transform inverses the sign of these coordinates, see \cite{Ser1} for the 
Schwarzschild's spacetime and the similar
description of the Reissner--Nordtr\"{o}m space-time in \cite{Chandr,Frolov}, for example.
Namely, for the two manifolds, A-manifold and B-manifold with coordinates $x$ and $\tilde{x}$, 
the symmetry $g_{\mu \nu}(x)\,=\,g_{\mu \nu}(\tilde{x})\,=\,\tilde{g}_{\mu \nu}(\tilde{x})$
must be preserved for a solution of Einstein equations by the CPTM symmetry transforms as following:
\beqar\label{Add1}
&\,&
q\,\rightarrow\,-\,\tilde{q}\,,r\,\rightarrow\,-\,\tilde{r}\,,
t\,\rightarrow\,-\,\tilde{t}\,,m_{grav}\,\rightarrow\,-\,\tilde{m}_{grav}\,;\,\,\,
\tilde{q}\,,
\,\tilde{r}\,,
\,\tilde{t}\,,
\,\tilde{m}_{grav}\,>\,0\,; \\
&\,&
CPTM(g_{\mu \nu}(x))\,=\,\tilde{g}_{\mu \nu}(\tilde{x})\,=\,g_{\mu \nu}(\tilde{x})\,\label{Add102}.
\eeqar
when the cosmological constant is zero. We underline, see \cite{Ser1}, 
that the usual radial coordinate is strictly positive and there is a need for an additional B-manifold in order to perform the 
\eq{Add1} discrete P transform, see also next Section. The transformation of the sign of the gravitational mass in this case can be understood as a consequence of the request of the preserving of the symmetry of metric.
Similar construction
in application of the quantum mechanics to the black hole physics and to Big Bang
and black hole singularities crossing  can be found in \cite{Hooft,Bars1} for example.

  The framework we consider consists of different manifolds with the gravitational masses of different signs in each one, see details in \cite{Ser1}.
The general motivation of the introduction of the negative mass in the different cosmological models is very clear. In any scenario, see 
\cite{Villata,Chardin} for example, the presence of some kind of the repulsive gravitation forces or an additional gravitational field in our Universe helps with an explanation of the existence of
dark energy in the models, see also 
\cite{DamKog,Petit1,Hoss,Kofinas,Fluid,Villata1,Hajd} and references therein.
It  is important, that the gravitation properties of the matter of B-manifold 
is also described by Einstein equations, see \cite{Villata,Villata1} or \cite{Chardin} and \cite{Souriau} for examples of the 
application of the discrete symmetries in the case of the quantum and classical systems. The situations becomes more complicated 
 when we notice the different directions of the
time's arrows in the different manifolds. Introducing the ordinary scalar fields, we obtain therefore a variant of the closed time path for the fields of the 
general manifold related by the CPTM symmetry with different time directions for the fields. The interaction between these fields leads to the quantum effective action calculated with respect to the fluctuations above the  classical solutions. Effectively, this effective action provides the cosmological constant in the classical Einsten's action. Namely, the vertices of the
effective  action "glues" the points of the same manifolds as well
as points of the different manifolds, allowing the gravitational interactions between them
that effectively reproduce the cosmological constant influence.
Once arising, the constant's value can be different for both manifolds related as well by the introduced symmetry. We will have then for the case of different cosmological constants for A and B manifolds:
\beq\label{Add103}
CPTM(\Lambda)\,=\,\tilde{\Lambda}\,,\,\,\,CPTM(g_{\mu \nu}(x, \Lambda))\,=\,\tilde{g}_{\mu \nu}(\tilde{x},\tilde{\Lambda})\,.
\eeq
This appearance of the cosmological constant satisfies the naturalness criteria of 't Hooft, see \cite{Hooft1}. It's zero value correspondence
to the precise symmetry between the metrics \eq{Add102} whereas it's small non-zero value decreases the symmetry to \eq{Add103} relation.
In general, the constant in the formalism is not a constant anymore but it is a functional which requires a renormalization depending on the form and properties of the interacting fields. It, therefore, can acquires different values due it's evolution, the important question we need to clarify in the formalism this is the present smallness of the constant.

 As we will see further, the mutual contribution of the scalar fields of two manifolds to the vacuum energy is zero at the classical level.
Respectively, there are two different mechanisms responsible for the constant's value which we will discuss through the paper. 
The first one is a dependence of the
cosmological constant on many loops quantum contribution to the vacuum energy. These contributions are provided by the ordinary framework of the quantum fields 
in the flat space-time, we will not discuss these contributions in the non-flat manifolds framework. In this case the smallness of the constant can be provided by the smallness of the corresponding diagrams in the QFT effective action or, alternatively, we demonstrate that the vacuum contributions
can be totally eliminated by the proper definition of the partition function of the theory.  An another contribution into the constant is a correction to the values of the vertices of the corresponding effective action due the 
non-flat corrections to the propagators and vertices. In this case the value of the constant is provided by the curvature corrections to the flat propagator and vertices of the scalar fields, in the almost flat space-time these correction are small. The additional separated question we investigate is about the bare value of the constant in the approach. The smallness of the bare 
cosmological constant, as we will demonstrate, can be provided by the characteristics of the scalar fields, there are different results for the massive and massless scalar fields for example.

 We consider a connection between the manifolds established through the effective action of the scalar fields defined on different manifolds. 
This effective action "glues" the manifolds, there is 
a kind of the foam of vertices that belongs to the same manifold as well as to both depending on the form and properties of the scalar fields.
In this case, the framework contains two or more manifolds which "talk" each with other by the  non-local correlators. 
These quantum vertices, similar to some extend to the quantum wormholes, have also been widely used in the investigation of the cosmological constant problem, see 
\cite{Fuller,ClassWorm,Wienberg} for example. 
The construction proposed here  is a dynamical one, the classical dynamics of metrics of separated manifolds can be affected by the quantum interactions between the manifolds. 
In any case, the cosmological constant arises in the
equations as a result of the mutual influence of the different parts of the extended manifold, i.e. due the worhmhole like interactions between the points of the same manifold or interactions between the
points of the different manifolds.
Solving the equations of motion  perturbatively, we begin from the case of the zero cosmological constant value in the case of "bare" separated manifolds and generate the constant at the next step of the evolution, breaking \eq{Add102} symmetry but preserving more general \eq{Add103} relations.
An interesting possibility, which we do not address in this paper, this is a possible large value of the cosmological constant at the 
beginning of the evolution and it's decreasing during the evolution. In this case, of course, the proposed perturbative scheme will not work. Nevertheless, at present the constant is small
and we can perform the calculations perturbatively.

 The appearance of the closed time path formalism in the framework and main properties of the CPTM transform are discussed in the Section \ref{S2}.
Sections \ref{S3} and \ref{S4} are dedicated to the formulation of the two different realization of CPTM symmetrical scalar fields in A and B manifolds
with overall zero classical contribution to the vacuum energy. In the Section \ref{S41} and Section \ref{S411}
 we describe the mechanism of the appearance of the cosmological constant
in the form of the interaction between the manifolds for the case of free fields. In the next Section we discuss the consequence of the symmetry on the form of the 
possible interactions terms between the fields of A and B manifolds. Sections \ref{S42} and \ref{S5} are about the inclusion of the interactions between the fields, 
in Section \ref{S5} we investigate an inclusion of the potentials of interaction and self-interacting of the fields considering as example usual $\phi^{4}$ scalar theory.
In Section \ref{S6} we consider the ways of the realization of the smallness of the constant in the formalism and Section \ref{S7} summarizes the obtained results.

\section{CPTM symmetry for scalar field }
\label{S2}

  In order to clarify the consequences of the CPTM symmetry, we will borrow some results from \cite{Ser1, Ser2}. 
As example of the symmetry application, we consider the Eddington--Finkelstein coordinates for the Schwarzschild spacetime
\beq\label{Sec1001}
\tv\,=\,t\,+\,r^{*}\,=\,t\,+\,r\,+\,2\, M\,\ln\left|\frac{r}{2\,M}\,-\,1\right|\,
\eeq
and
\beq\label{Sec1002}
\tu\,=\,t\,-\,r^{*}\,.
\eeq
In correspondence, we define also the
Kruskal--Szekeres coordinates $U,V$, which covers the whole extended space-time, are defined in the different parts of the extended solution.
For example, when considering  the region \rom{1} with $r\,>\,2M$ in terms of \cite{Frolov} 
where $U\,<\,0\,,V\,>\,0$ we have: 
\beq\label{Sec1003}
U\,=\,-\,e^{-\tu / 4M}\,,\,\,\,\,V\,=\,e^{\tv / 4M}\,.
\eeq
As demonstrated in \cite{Ser1}, see also \cite{Frolov}, the transition to the separated regions of the solutions can by done by the analytical continuation of the coordinates provided by the 
corresponding change of its signs and reversing of the sign of the gravitational mass. When considering the region \rom{3} 
in definitions of~\cite{Frolov}, we obtain:
\beqar\label{Sec8}
U\,=\,-\,e^{-\tu / 4M}\,\rightarrow\,\tilde{U}\,=\,e^{-\tilde{\tu} / 4\tilde{M}}\,=\,-\,U\,,\\
V\,=\,e^{\tv / 4M}\,\rightarrow\,\tilde{V}\,=\,-\,e^{-\tilde{\tv} / 4\tilde{M}}\,=\,-\,V\,.
\eeqar
This inversion of the signs of the $(U,V)$ coordinate axes will hold correspondingly in the all regions of $(U,V)$ plane after 
analytical continuation introduced in \cite{Ser1}.   
Formally, from the point of view of the discrete transform performed in $(U,V)$ plane, the transformations \eq{Sec8}
are equivalent to the full reversion of the Kruskal-Szekeres "time" 
\beq\label{Sec1004}
T\,=\,\frac{1}{2}\,\Le\,V\,+\,U\,\Ra\,\rightarrow\,-\,T
\eeq
and radial "coordinate"
\beq\label{Sec1005}
R\,=\,\frac{1}{2}\,\Le\,V\,-\,U\,\Ra\,\rightarrow\,-\,R
\eeq
in the complete Schwarzschild space-time. Therefore, the introduced $T,R$ coordinates and some
transverse coordinates $X_{\bot}$, all denoted simply as $x$,
can be considered as the correct coordinates in definition of the quantum fields.  The corresponding change of the coordinates
in the expressions of the functions after the Fourier transform, which relates the fields from the different manifolds, 
being formally similar to the conjugation is not the conjugation.
Namely, the analytical continuation of the functions from A-manifold to B-manifold (CPTM transform) means the change of the sign of $x$ in 
corresponding Fourier expressions without its conjugation as whole.

 Coming back to the usual coordinates in the each A or B manifolds separately, we will obtain for the radial coordinate:
\beqar\label{Dop1}
\int_{0}^{\infty}\,d R\rightarrow\,\int_{0}^{\infty}\,r\,,\,\,\,\,\,\,R\rightarrow\,r 
\\
\int_{-\infty}^{0}\,d R\rightarrow\,\int_{0}^{\infty}\,\tilde{r}\,,\,\,\,\,\,\,r\rightarrow\,-\,\tilde{r}\,, 
\eeqar
and time:
\beqar\label{Dop2}
\int_{-\infty}^{\infty}\,d T\rightarrow\,\int_{-\infty}^{\infty}\,t\,,\,\,\,\,\,\,T\rightarrow\,t 
\\
\int_{-\infty}^{\infty}\,d T\rightarrow\,\int_{-\infty}^{\infty}\,\tilde{t}\,,\,\,\,\,\,\,t\rightarrow\,-\,\tilde{t}\,. 
\eeqar
\begin{figure}[!hb]
\centering
\psfig{file=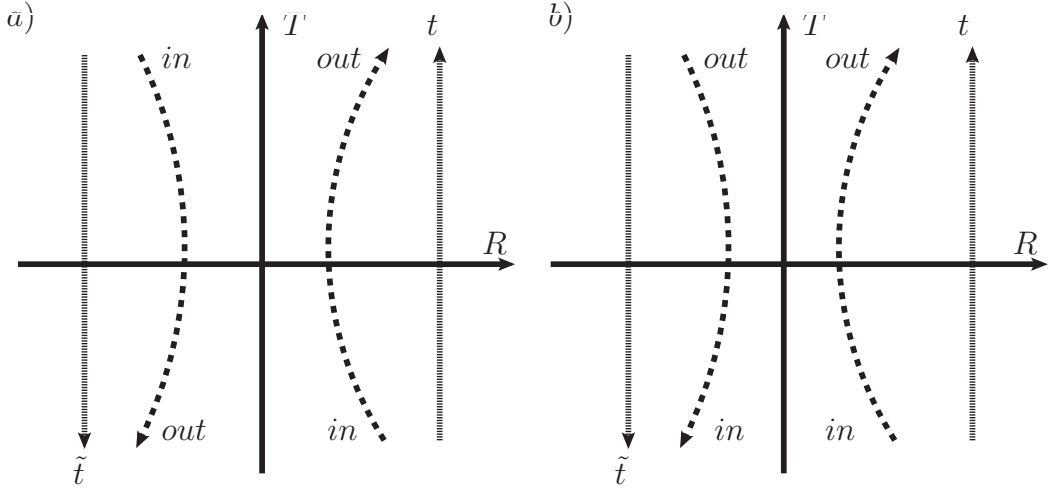,width=140mm} 
\caption{The diagram represents two different possibilities of the construction of the closed time paths.}
\label{Pic1}
\end{figure}
There are, therefore, two possibilities to determine the time's arrows in both manifolds depending on the determination of $in$ and $out$ states for the second manifold, see \fig{Pic1},
correspondingly there are two possibilities for the definition of the form of the scalar field of B manifold.
Namely, for the application of the introduced symmetry, we consider A and B manifolds (two Minkowski spaces) as separated parts of the extended solution with non-interacting branches of the scalar quantum field 
defined in each region and related by the CPTM discrete transform.
Now consider the usual quantum scalar field defined in our part (A-manifold) of the extended solution:
\beq\label{Sec1006}
\phi(x)\,=\,\int\,\frac{d^{3}\,k}{(2\pi)^{3/2}\,\sqrt{2\, \omega_{k}}}\,\Le
\phi^{-}(k)\,e^{-\imath\,k\,x}\,+\,\phi^{+}(k)\,e^{\imath\,k\,x}\Ra\,=\, \phi^{*}(x) \,,\,\,\,\,[\phi^{-}(k),\,\phi^{+}(k^{'})]\,=\,\delta^{3}_{k\,k^{'}}\,.
\eeq
The conjugation of the scalar field does not change the expressions, we have simply $(\phi^{-})^{*}\,=\,\phi^{+}$.
In contrast to the conjugation, the CPTM discrete transform acts differently. We have for the second field:
\beqar\label{Sec1}
CPTM(\phi(x))\, & =  &\,CPTM\,\Le\,\int\,\frac{d^{3}\,k}{(2\pi)^{3/2}\,\sqrt{2\,\omega_{k}}}\,\Le
\phi^{-}(k)\,e^{-\imath\,k\,x}\,+\,\phi^{+}(k)(k)\,e^{\imath\,k\,x}\Ra\,\Ra\,\,=\, 
\nonumber \\
&=&\,
\tilde{\phi}(\tilde{x})\,=\,\int\,\frac{d^{3}\,k}{(2\pi)^{3/2}\,\sqrt{2\,\tilde{\omega}_{k}}}\,\Le
\phi^{-}(k)\,e^{\imath\,k\,\tilde{x}}\,+\,\phi^{+}(k)(k)\,e^{-\imath\,k\,\tilde{x}}\Ra\,.
\eeqar
Depending on the commutation relations between the components of the field of B manifold, we will obtain two different types of the scalar field in correspondence to 
\fig{Pic1}. This issue we discuss in the following two Sections considering the fields in flat A,B manifolds.

\section{Scalar field of B-manifold: first possibility }
\label{S3}

 As mentioned above, there are two possibilities for the definition of the B-scalar fields which we consider. We begin from the obvious one, defining the B-field similarly
to the definition of the usual scalar field with the vacuum state mutual for both manifolds:
\beqar\label{Sec199}
CPTM(\phi(x))\,& = &\,CPTM\,\Le\,\int\,\frac{d^{3}\,k}{(2\pi)^{3/2}\,\sqrt{2\,\omega_{k}}}\,\Le
\phi^{-}(k)\,e^{-\imath\,k\,x}\,+\,\phi^{+}(k)\,e^{\imath\,k\,x}\Ra\,\Ra\,\,=\, 
\nonumber \\
&=&\,\tilde{\phi}(\tilde{x})\,=\,
\int\,\frac{d^{3}\,k}{(2\pi)^{3/2}\,\sqrt{2\,\tilde{\omega}_{k}}}\,\Le \tilde{\phi}^{+}(k)\,e^{\imath\,k\,\tilde{x}}\,+\,
\tilde{\phi}^{-}(k)\,e^{-\imath\,k\,\tilde{x}}\,\Ra\,
\eeqar
with the following properties of the annihilation and creation operators of B-scalar field:
\beq\label{Sec101}
\left\{ 
\begin{array}{c}
\phi^{-}(k)\,\leftrightarrow\,\tilde{\phi}^{+}(k)\,\\
\phi^{+}(k)\,\leftrightarrow\,\tilde{\phi}^{-}(k)\,
\end{array}
\right.\,\rightarrow\,
\left\{ 
\begin{array}{c}
[\,\tilde{\phi}^{-}(k)\,\tilde{\phi}^{+}(k^{'})\,]\,=\,-\,\delta^{3}_{k\,k^{'}}\,\\
<\,\tilde{\phi}^{-}(k)\,\tilde{\phi}^{+}(k^{'})\,>\,=\,-\,\delta^{3}_{k\,k^{'}}\,
\end{array}
\right.\,
\eeq
and
\beq\label{Sec2}
\omega(k)\,=\,\sqrt{m^2 + k^2}\rightarrow\,\tilde{\omega}(k)\,=\,\sqrt{m^2 + k^2}\,.
\eeq
Still, the definitions above are meaningless if we do not define the vacuum states for the A and B scalar fields. There is a
mutual vacuum state in our problem with the following properties:
\beq\label{VS3}
\left\{ 
\begin{array}{c}
<0|\,\phi^{+}\,=\,0\,\\
\phi^{+}\,|0>\neq\,0\,
\end{array}
\right.\,\stackrel{CPTM}{\leftrightarrow}
\left\{ 
\begin{array}{c}
\tilde{\phi}^{-}\,|0>=\,0\,\\
<0|\,\tilde{\phi}^{-}\,\neq\,0\,
\end{array}
\right.;\,\,\,
\left\{ 
\begin{array}{c}
<0|\,\phi^{-}\,\neq\,0\,\\
\phi^{-}\,|0>\,=\,0\,
\end{array}
\right. \,\,\stackrel{CPTM}{\leftrightarrow}
\left\{ 
\begin{array}{c}
\tilde{\phi}^{+}\,|0>\,\neq\,0\,\\
<0|\,\tilde{\phi}^{+}\,=\,0\,
\end{array}
\right. \,.
\eeq
A general energy-momentum vector $P_{\mu}$ written for the both regions of the extended manifold and averaged with respect to the mutual vacuum state
now acquires the following form:
\beqar\label{Sec4}
<P^{\mu}> & = &\frac{1}{2}\,\int\,d^{3}\,k\,k^{\,\mu} \Le <\Le \phi^{+}(k)\,\phi^{-}(k)+
\phi^{-}(k)\,\phi^{+}(k)\Ra >  + 
\right.
\nonumber \\
&+&
\left.
<\Le \tilde{\phi}^{+}(k)\,\tilde{\phi}^{-}(k)\,+\,\tilde{\phi}^{-}(k)\,\tilde{\phi}^{+}(k)\Ra |>\Ra\,=\,
\nonumber \\
&=&
\int\,d^{3}\,k\,k^{\,\mu} \Le <\phi^{+}(k)\,\phi^{-}(k)>  + 
<\tilde{\phi}^{+}(k)\,\tilde{\phi}^{-}(k)>\Ra\,=\,
\nonumber \\
&=&
<P^{\mu}_{A}>\,+\,<P^{\mu}_{B}>\,=\,0\,.
\eeqar
Here $P^{\mu}_{A,B}$ are the energy-momentum vectors of A or B manifolds separately, 
we also note that
\beq\label{Sec41}
CPTM(<P^{\mu}_{A}>)\,=\,<P^{\mu}_{B}>\,
\eeq
as expected.
So, as a consequence of CPTM symmetry, we obtained the precise cancellation of the vacuum zero modes contributions. 

 The next issue we discuss in connection to the B-field is a definition of the propagators of the field. For the easier references we presented the forms of the Feynman and Dyson propagators we used in the Appendixes A. In general we need to determine the change of the propagators in respect to the CPTM transform
and form of the propagator of the B-field which we will use in the calculations.
Therefore, we consider the usual $\tilde{G}_{F}(\tilde{x}-\tilde{y})$ propagator for the B-field, it has the following form:
\beq\label{Sec5}
\tilde{G}_{F}(\tilde{x},\tilde{y})\,= \,
-\imath\,\Le \theta(\tilde{x}^{0}-\tilde{y}^{0})<\tilde{\phi}(\tilde{x})\,\tilde{\phi}(\tilde{y})>\,+\,
\theta(\tilde{y}^{0}-\tilde{x}^{0})<\tilde{\phi}(\tilde{y})\,\tilde{\phi}(\tilde{x})> \Ra\,=\,-\,G_{F}(\tilde{x},\tilde{y}),
\eeq
see \eq{Sec101} definitions. Alternatively we can calculate
\beqar\label{Sec6}
CPTM(G_{F}(x,y))& = &-\imath\Le \theta(\tilde{y}^{0}-\tilde{x}^{0}) CPTM(<\phi(x)\,\phi(y)>)+\theta(\tilde{x}^{0}-\tilde{y}^{0})
CPTM(<\phi(y)\,\phi(x)>)\Ra =
\nonumber \\
&=&
\imath\Le \theta(\tilde{y}^{0}-\tilde{x}^{0})\,<\phi(\tilde{y})\,\phi(\tilde{x})>+
\theta(\tilde{x}^{0}-\tilde{y}^{0})\,<\phi(\tilde{x})\,\phi(\tilde{y})>\Ra=-\,G_{F}(\tilde{x},\tilde{y})\,,
\eeqar
that coincides with \eq{Sec5} answer. In \eq{Sec6} the following property of Wightman function under the CPTM transform is clarified:
\beq\label{Sec7}
CPTM(D(x-y))\,\propto \,CPTM\Le\,<\ph^{-}(k)\,\ph^{+}(k^{'})>\,e^{-\imath\,k\,x\,+\,\imath\,k^{'}\,y}\,\Ra\,=\,
<\tilde{\ph}^{-}(k^{'})\,\tilde{\ph}^{+}(k)>\,e^{\imath\,k\,\tilde{x}\,-\,\imath\,k^{'}\,\tilde{y}}\,
\eeq
that provides
\beq\label{Sec8801}
CPTM(D(x-y))\,=\,
-\,D(\tilde{y}\,-\,\tilde{x})\,=\,-\,<\phi(\tilde{y})\,\phi(\tilde{x})>\,
\eeq
in accordance to \eq{Sec6}.

\section{Scalar field of B-manifold: second possibility }
\label{S4}

  An another way do define the B-field is to consider it as an antimatter field of A-field in respect to the sign of the mass. We define therefore:
\beqar\label{VS1}
CPTM(\phi(x))\,& = &\,CPTM\,\Le\,\int\,\frac{d^{3}\,k}{(2\pi)^{3/2}\,\sqrt{2\,\omega_{k}}}\,\Le
\phi^{-}(k)\,e^{-\imath\,k\,x}\,+\,\phi^{+}(k)\,e^{\imath\,k\,x}\Ra\,\Ra\,\,=\, 
\nonumber \\
&=&\,\tilde{\phi}(\tilde{x})\,=\,
\int\,\frac{d^{3}\,k}{(2\pi)^{3/2}\,\sqrt{2\,\tilde{\omega}_{k}}}\,\Le
\tilde{\phi}^{-}(k)\,e^{\imath\,k\,\tilde{x}}\,+\,\tilde{\phi}^{+}(k)\,e^{-\imath\,k\,\tilde{x}}\Ra\,
\eeqar
with the following properties of the operators:
\beq\label{VS2}
\left\{ 
\begin{array}{c}
\phi^{-}(k)\,\leftrightarrow\,\tilde{\phi}^{-}(k)\,\\
\phi^{+}(k)\,\leftrightarrow\,\tilde{\phi}^{+}(k)\,
\end{array}
\right.\,\rightarrow\,
\left\{ 
\begin{array}{c}
[\,\tilde{\phi}^{-}(k)\,\tilde{\phi}^{+}(k^{'})\,]\,=\,\delta^{3}_{k\,k^{'}}\,\\
<\,\tilde{\phi}^{+}(k)\,\tilde{\phi}^{-}(k^{'})\,>\,=\,-\,\delta^{3}_{k\,k^{'}}\,
\end{array}
\right.\,.
\eeq
Correspondingly we define the action of these operators on the vacuum state as following:
\beq\label{VS3301}
\left\{ 
\begin{array}{c}
<0|\,\phi^{+}\,=\,0\,\\
\phi^{+}\,|0>\neq\,0\,
\end{array}
\right.\,\stackrel{CPTM}{\leftrightarrow}
\left\{ 
\begin{array}{c}
\tilde{\phi}^{+}\,|0>=\,0\,\\
<0|\,\tilde{\phi}^{+}\,\neq\,0\,
\end{array}
\right.;\,\,\,
\left\{ 
\begin{array}{c}
<0|\,\phi^{-}\,\neq\,0\,\\
\phi^{-}\,|0>\,=\,0\,
\end{array}
\right. \,\,\stackrel{CPTM}{\leftrightarrow}
\left\{ 
\begin{array}{c}
\tilde{\phi}^{-}\,|0>\,\neq\,0\,\\
<0|\,\tilde{\phi}^{-}\,=\,0\,
\end{array}
\right. \,.
\eeq
In this case the general energy-momentum vector $P_{\mu}$ written for the both regions of the extended manifold again is defined as usual and we obtain:
\beqar\label{VS4}
<P^{\mu}> & = &\frac{1}{2}\,\int\,d^{3}\,k\,k^{\,\mu} \Le <\,\Le \phi^{+}(k)\,\phi^{-}(k)+
\phi^{-}(k)\,\phi^{+}(k)\Ra\,>  + 
\right.
\nonumber \\
&+&
\left.
<\,\Le \tilde{\phi}^{+}(k)\,\tilde{\phi}^{-}(k)\,+\,\tilde{\phi}^{-}(k)\,\tilde{\phi}^{+}(k)\Ra \,>\Ra\,=\,
\nonumber \\
&=&\,
\int\,d^{3}\,k\,k^{\,\mu} \Le <\,\phi^{+}(k)\,\phi^{-}(k)\,>  + 
<\,\tilde{\phi}^{-}(k)\,\tilde{\phi}^{+}(k)\,>\Ra\,=\,
\nonumber \\
&=&
<P^{\mu}_{A}>\,+\,<P^{\mu}_{B}>\,=\,
0\,
\eeqar 
with
\beq\label{VS5}
CPTM(<P^{\mu}_{A}>)\,=\,<P^{\mu}_{B}>\,.
\eeq
as above.

 Now we once more define the 
$\tilde{G}_{F}(\tilde{x}-\tilde{y})$ propagator for the B-field, it has the following form:
\beq\label{VS6}
\tilde{G}_{F}(\tilde{x},\tilde{y})\,= \,
-\imath\,\Le \theta(\tilde{x}^{0}-\tilde{y}^{0})<\tilde{\phi}(\tilde{x})\,\tilde{\phi}(\tilde{y})>\,+\,
\theta(\tilde{y}^{0}-\tilde{x}^{0})<\tilde{\phi}(\tilde{y})\,\tilde{\phi}(\tilde{x})> \Ra\,=\,-\,G_{D}(\tilde{x},\tilde{y})
\eeq
which is different from the \eq{Sec5} expression due the \eq{VS2} properties of the operators. Checking the CPTM symmetry connection of the A and B fields we have:
\beqar\label{VS7}
CPTM(G_{F}(x,y))& = &-\imath\Le \theta(\tilde{y}^{0}-\tilde{x}^{0}) CPTM(<\phi(x)\,\phi(y)>)+\theta(\tilde{x}^{0}-\tilde{y}^{0})
CPTM(<\phi(y)\,\phi(x)>)\Ra =
\nonumber \\
&=&
\imath\Le \theta(\tilde{y}^{0}-\tilde{x}^{0})\,<\phi(\tilde{x})\,\phi(\tilde{y})>+
\theta(\tilde{x}^{0}-\tilde{y}^{0})\,<\phi(\tilde{y})\,\phi(\tilde{x})>\Ra=-\,G_{D}(\tilde{x},\tilde{y})\,
\eeqar
as expected\footnote{We notice, that in both cases the kinetic term for the B scalar field is negative. Nevertheless, these fields are not ghost fields widely used
in the theories of dark matter. Our A and B fields are present on the separated manifolds and negative sign of the B field is due a choice 
of the positive direction for the time and radial coordinates, we juxtapose everything with the directions in our chosen A manifold. }.
Again, we can understand this transform as consequence of the transformation of the 
Wightman function under the CPTM transform:
\beq\label{VS8}
CPTM(D(x-y))\,\propto \,CPTM\Le\,<\ph^{+}(k)\,\ph^{-}(k^{'})>\,e^{\imath\,k\,x\,-\,\imath\,k^{'}\,y}\,\Ra\,=\,
<\tilde{\ph}^{+}(k)\,\tilde{\ph}^{-}(k^{'})>\,e^{-\imath\,k\,\tilde{x}\,+\,\imath\,k^{'}\,\tilde{y}}\,
\eeq
that provides
\beq\label{vs9}
CPTM(D(x-y))\,=\,
-\,D(\tilde{x}\,-\,\tilde{y})\,=\,-\,<\phi(\tilde{x})\,\phi(\tilde{y})>\,.
\eeq
in correspondence to \eq{VS6}-\eq{VS7} result.

\section{Action of the formalism: free fields with single source}
\label{S41}
 
 Introducing the action for the scalar fields, we use the formalism proposed in \cite{Ser2} with the full action defined at the absence of the matter as 
\beq\label{CC1}
S\,=\,S_{grav}(x,\,\tilde{x})\,+\,S_{int}(x,\,\tilde{x})\,,
\eeq
where
\beq\label{CC2}
S_{grav}(x,\,\tilde{x})\,=\,-\,m_{p}^{2}\,\int_{-\infty}^{\infty} dt\int\,d^{3}x\,\sqrt{-g(x)}\,R(x)\,-\,
\,m_{p}^{2}\,\int_{-\infty}^{\infty} d\tilde{t}\int\,d^{3}\tilde{x}\,\sqrt{-\tilde{g}(\tilde{x})}\,R(\tilde{x})\,
\eeq
are separated Einstein actions defined in each A-B manifolds separately
and
\beq\label{CC3}
S_{int}\,=\,-\,\sum_{i,j=A,B}\,m_{p}^{2}\,\int\,d^4 x_{i}\,\sqrt{-g_{i}(x_{i})}\,\int d^4 x_{j}\sqrt{-g(x_{j})}\,\xi_{i j}(x_{i}, x_{j})\,
\eeq
is a simplest terms which describes the gravitational interactions  between the manifolds through some bi-scalar functions $\xi_{i j}(x_{i}, x_{j})$, here $x_{A}\,=\,x$ and
$x_{B}\,=\,\tilde{x}$ are defined and
 $m_{p}^{2}\,=\,1\,/\,16\,\pi\,G$ with $c\,=\,\hbar\,=\,1$ notations are introduced for simplicity.
Without this term the system is a system for non-coupled Einstein equations for each manifold separately, whereas the interaction term in the equations determines the mutual 
influence of the manifolds in the form of the cosmological constant. 

 In order to understand the appearance of the interaction term in \eq{CC1} we, following to \cite{Ser2}, consider 
the discussed above 
scalar fields in A-B manifolds with some physical sources of them introduced.
Now, instead \eq{CC3}, we consider  the action for the free non-interacting scalar fields in a curved space-time:
\beq\label{CC4}
S_{int}\,=\,\int\, d^4 x\,\sqrt{-g}\,\Le \,\frac{1}{2}\,\phi\,G^{-1} \,\phi \,-\,m_{p}^{3}\,f(x)\,\phi\,\Ra\,+\,
\int\, d^4 \tilde{x}\,\sqrt{-\tilde{g}}\,\Le\,\frac{1}{2}\, \tilde{\phi}\,\tilde{G}^{-1} \,\tilde{x}\,\tilde{\phi} \,\mp\,m_{p}^{3}\,
\tilde{f}(\tilde{x})\,\tilde{\phi}\,\Ra\,.
\eeq
Here we use the notations for the propagators in the curved space-time, see Appendix B. 
We also do not know a priori the sign for the second source in the 
Lagrangian\footnote{We introduced in \eq{CC4} the undefined function $f(x)\,=\,\tilde{f}(\tilde{x})$ as a source of the scalar field. Further, 
where the form of the function does not matter,
we will take it equal to 1.}, namely it is possible that $CPTM(f)\,=\,\pm\,\tilde{f}$, 
and further, where it will be need in that, will consider the two cases separately.
Now the both fields can be expanded around their classical values:
\beq\label{CC5}
\phi\,=\,m_{p}^{3}\,G(x,y,\Lambda)\,f(y)\,\sqrt{-g(y,\Lambda)}\,+\,\varepsilon\,,\,\,\,
\tilde{\phi}\,=\,\pm\,m_{p}^{3}\,\tilde{G}(\tilde{x},\tilde{y},\tilde{\Lambda})\,\tilde{f}(\tilde{y})\,\sqrt{-\tilde{g}(\tilde{y},\tilde{\Lambda})}\,+\,
\tilde{\varepsilon}\,
\eeq
that provides for the action in both cases:
\beqar\label{CC6}
S_{int}\,& = &\,\frac{1}{2}\,\int\, d^4 x\,\sqrt{-g}\,\varepsilon\,G^{-1} \,\varepsilon \,-\,\frac{m_{p}^{6}}{2}\,\int\, d^4 x\,\int\, d^4 y\,\sqrt{-g(x)}\,f(x)\,
G(x,y)\,f(y)\,\sqrt{-g(y)}\,+
\nonumber \\
&+&
\,\frac{1}{2}\,\int\, d^4 \tilde{x}\,\sqrt{-g}\,\tilde{\varepsilon}\,\tilde{G}^{-1} \,\tilde{\varepsilon} \,-\,
\frac{m_{p}^{6}}{2}\,\int\, d^4 \tilde{x}\,\int\, d^4 \tilde{y}\,\sqrt{-\tilde{g}(\tilde{x})}\,\tilde{f}(\tilde{x})\,\tilde{G}(\tilde{x},\tilde{y})\,
\tilde{f}(\tilde{y})\,
\sqrt{-\tilde{g}(\tilde{y})}\,,
\eeqar
for simplicity we did not write here and further an implicit  dependence of the metric on the manifold's cosmological constants.
We obtain, therefore, that the first and third terms in the expression can be integrated out, whereas the second and fourth terms in the action determine the cosmological constant in the expressions through the $G_{F}(x,y)$ and $\tilde{G}_{F}(\tilde{x},\tilde{y})$ propagators after the integration over $y$ and 
$\tilde{y}$ coordinates for the A and B manifolds correspondingly. 


The cosmological constant in this case contributes to the usual equations of motion:
\beqar\label{CC501}
\delta\,S_{\Lambda\,\tilde{\Lambda}}& = &
\frac{m_{p}^{6}}{2}\int d^4 x\int d^4 y\,\sqrt{-g(x)}\,f(x)\,\Le  g_{\mu \nu} \delta g^{\mu \nu}\Ra G(x,y)\,f(y)\,\sqrt{-g(y)}\,-
\nonumber \\
&-&
\frac{m_{p}^{6}}{2}\int d^4 x\int\, d^4 y\,\sqrt{-g(x)}\,f(x)\,\Le \delta\,G(x,y)\,f(y)\,\Ra\sqrt{-g(y)}\,
\,+\,
\nonumber \\
&+&
\frac{m_{p}^{6}}{2}\int d^4 \tilde{x}\int d^4 \tilde{y}\,\sqrt{-\tilde{g}(\tilde{x})}\,\tilde{f}(\tilde{x})\,
\Le  
\tilde{g}_{\mu \nu} \delta \tilde{g}^{\mu \nu}\Ra \tilde{G}(\tilde{x},\tilde{y})\,\tilde{f}(\tilde{y})\,\sqrt{-\tilde{g}(\tilde{y})}\,-
\nonumber \\
&-&
\frac{m_{p}^{6}}{2}\int d^4 \tilde{x}\int\, d^4 \tilde{y}\,\sqrt{-\tilde{g}(\tilde{x})}\,\tilde{f}(\tilde{x})\,
\Le \delta\,\tilde{G}(\tilde{x},\tilde{y})\Ra\,\tilde{f}(\tilde{y})\,
\sqrt{-\tilde{g}(\tilde{y})}\,=\,
\nonumber \\
& = &\,m_{p}^{2}\int d^4 x\,\sqrt{-g(x)}\Le  g_{\mu \nu} \delta g^{\mu \nu}\Ra\,\Lambda(x)\,-\,
2\,m_{p}^{2}\int d^4 x\,\sqrt{-g(x)}\,\delta\,\Lambda(x)\,+\,
\nonumber \\
&+&\,
\,m_{p}^{2}\,\int d^4 \tilde{x}\,\sqrt{-\tilde{g}(\tilde{x})}\Le  \tilde{g}_{\mu \nu} \delta \tilde{g}^{\mu \nu}\Ra\, \tilde{\Lambda}(\tilde{x})\,-\,
2\,m_{p}^{2}\,\int d^4 \tilde{x}\,\sqrt{-\tilde{g}(\tilde{x})}\,\delta\, \tilde{\Lambda}(\tilde{x})\,
\eeqar
which is correct when we take zero cosmological constant in the metrics in r.h.s. of \eq{CC6}, see Appendix B. Here, in \eq{CC501}, the propagators must be understood as the full ones in the curved space-time, again see Appendix B definitions.  In general, if we want a precise equation for the cosmological constants, then 
we have also to account a presence of them in the metric and or to resolve the equations as whole, obtaining a non-perturbative solution, or use 
\eq{B13} prescription obtaining perturbative solution for the constants. To the first approximation, we obtain:
\beqar\label{CC8}
&\,& \Lambda(x)\,+\,\tilde{\Lambda}(\tilde{x})\,= \,
\nonumber \\
& = &
\frac{m_{p}^{4}}{4} \,\Le
\,f(x)\,\int\, d^4 y\,G(x,y,\Lambda)\,f(y)\,\sqrt{-g(y,\Lambda)}\,+
\,\tilde{f}(\tilde{x})\,\int d^4 \tilde{y}\,\tilde{G}(\tilde{x},\tilde{y},\tilde{\Lambda})\,\tilde{f}(\tilde{y})\,
\sqrt{-\tilde{g}(\tilde{y},\tilde{\Lambda})} \Ra\,=\,
\nonumber \\
&=&
\frac{m_{p}^{4}}{4} \,\Le
\,f(x)\,\int\, d^4 y\,G(x,y)\,f(y)\,\sqrt{-g(y)}\,+
\,\tilde{f}(\tilde{x})\,\int d^4 \tilde{y}\,\tilde{G}(\tilde{x},\tilde{y})\,\tilde{f}(\tilde{y})\,
\sqrt{-\tilde{g}(\tilde{y})} \Ra\,
\eeqar
where as propagators we take \eq{B6} expression for the full propagators with corresponding definition of "bare" propagators for A and B manifolds. 

 For the case of the first type of scalar fields, 
preserving in \eq{CC8} the leading terms from \eq{B6} with respect to the flat propagators, we obtain:
\beqar\label{CC801}
\Lambda(x)\,&=& \,\frac{m_{p}^{4}}{4} \,f(x)\,\int\, d^4 y\, \sqrt{-g(y)}\,G(x,y)\,f(y)\,=\,
\frac{m_{p}^{4}}{4} \,f(x)\,\int\, d^4 y\, \sqrt{-g(y)} \,G_{F}(x,y)\,f(y)\,
\nonumber \\
\tilde{\Lambda}(x)\,& = &\,\frac{m_{p}^{4}}{4} \,\tilde{f}(x)\,\int\, d^4 y\, \sqrt{-g(y)}\,\tilde{G}(x,y)\,\tilde{f}(y)=
-\frac{m_{p}^{4}}{4} \,\tilde{f}(x)\,\int\, d^4 y\, \sqrt{-g(y)}\,G_{F}(x,y)\,\tilde{f}(y)\,.
\eeqar
We have therefore to this order of precision
\beq\label{CC802}
\Lambda\,=\,-\,\tilde{\Lambda}\,.
\eeq
Nevertheless, effectively, in the Lagrangian, we have for these terms
\beq\label{CC8020}
\Lambda\,+\,\tilde{\Lambda}\,=\,0\,
\eeq
and the constants do not appear in the action in this perturbative order. The first non-trivial contribution in the Lagrangian, therefore, will
be given by the quadratic with respect to the flat propagators terms in \eq{B6} expression.

In the case of the second type of the fields, also keeping in the \eq{CC8} only the first non-vanishing terms from \eq{B6}, we obtain:
\beqar\label{CC8031}
\Lambda(x)\,& = &\,\frac{m_{p}^{4}}{4}\,f(x)\, \int\, d^4 y\,\sqrt{-g(y)} \,G_{F}(x,y)\,f(y)\,
\nonumber \\
\tilde{\Lambda}(x)\, & = &\,-\,\frac{m_{p}^{4}}{4} \,\tilde{f}(\tilde{x})\,\int\, d^4 y\, \sqrt{-g(y)} \,G_{D}(x,y)\,\tilde{f}(\tilde{y})\,
\eeqar
that provides
\beq\label{CC804}
\Lambda(x)\,=\,\tilde{\Lambda}^{*}(x)\,,
\eeq
see again Appendix A for the definitions of the propagators in the flat space-time.

\section{Action of the formalism: free fields with double sources}
\label{S411}

 In the previous section we consider the simplest Lagrangian  of the free scalar fields, there is no interacting between them in the Lagrangian.
Further, due the closed time path introduced in the action, such term will arise correspondingly to the Keldysh formalism prescription. Nevertheless, 
in the Keldysh approach, the contribution of this kinetic term in the effective action begins from one loop of the corresponding propagator  and, therefore, it's contribution is small in general. So, as we obtained in the previous Section, see \eq{CC6}, in the previous case there are no interactions between the A and B manifolds but only self-interactions between the points of the same manifolds. 
In order to introduce the interactions, in this Section we consider a more complicated version of \eq{CC4} with
two sources for the scalar fields present as following:
\beqar\label{DS1}
S_{int}\,& = &\,\int\, d^4 x\,\sqrt{-g}\,\Le \,\frac{1}{2}\,\phi\,G^{-1} \,\phi \,-\,m_{p}^{3}\,f(x)\,\phi\,-\,
\xi\,m_{p}^{3}\,f(x)\,\tilde{\phi}\,\Ra\,+\,
\nonumber \\
&+&\,
\int\, d^4 \tilde{x}\,\sqrt{-\tilde{g}}\,\Le\,\frac{1}{2}\, \tilde{\phi}\,\tilde{G}^{-1} \,\tilde{x}\,\tilde{\phi} \,-\,
\tilde{\xi}\,m_{p}^{3}\,\tilde{f}(\tilde{x})\,\tilde{\phi}\,-\,\xi\,\tilde{\xi}\,m_{p}^{3}\,\tilde{f}(\tilde{x})\,\phi\,\Ra\,,
\eeqar
the parameters $\xi\,=\,\pm\,1$ and $\tilde{\xi}=\pm\,1$ are defined here independently and
the sign of $f$ is undefined, it can be positive or negative, whereas the sign of $\tilde{f}$ is strictly positive.
We solve the equations of motion and obtain the classical solutions for the fields: 
\beqar\label{DS2}
\phi\,& = &\,m_{p}^{3}\,G(x,y,\Lambda)\,\Le f(y)\sqrt{-g(y,\Lambda)}\,
 + \xi\,\tilde{\xi}\,\tilde{f}(y))\sqrt{-\tilde{g}(y,\tilde{\Lambda})}\,\Ra\,+\,\varepsilon\,=\,
\phi_{cl}\,+\,\varepsilon\,,
\\
\tilde{\phi}\,& = &\,m_{p}^{3}\,\tilde{G}(\tilde{x},\tilde{y},\tilde{\Lambda})\,
\Le \tilde{\xi}\,\tilde{f}(\tilde{y})\sqrt{-\tilde{g}(\tilde{y},\tilde{\Lambda})}\,
 + \xi\,f(\tilde{y})\sqrt{-g(\tilde{y},\Lambda)}\,\Ra\,
+\,\tilde{\varepsilon}\,=\,\tilde{\phi}_{cl}\,+\,\tilde{\varepsilon}\,.
\eeqar
Performing the usual calculations we will obtain for the \eq{DS1} expression: 
\beqar\label{DS3}
S_{int}\,& = &\,\frac{1}{2}\,\int\, d^4 x\,\sqrt{-g}\,\varepsilon\,G^{-1} \,\varepsilon \,-\,\frac{m_{p}^{6}}{2}\,\int\, d^4 x\,\int\, d^4 y\,\sqrt{-g(x)}\,f(x)\,
G(x,y)\,f(y)\,\sqrt{-g(y)}\,-
\nonumber \\
&-&
\,\frac{m_{p}^{6}}{2}\,\int\, d^4 x\,\int\, d^4 y\,\sqrt{-\tilde{g}(x)}\,\tilde{f}(x)\,G(x,y)\,\tilde{f}(y)\,\sqrt{-\tilde{g}(y)}\,+
\nonumber \\
&+&
\,\frac{1}{2}\,\int\, d^4 \tilde{x}\,\sqrt{-g}\,\tilde{\varepsilon}\,\tilde{G}^{-1} \,\tilde{\varepsilon} \,-\,
\frac{m_{p}^{6}}{2}\,\int\, d^4 \tilde{x}\,\int\, d^4 \tilde{y}\,\sqrt{-\tilde{g}(\tilde{x})}\,\tilde{f}(\tilde{x})\,\tilde{G}(\tilde{x},\tilde{y})\,
\tilde{f}(\tilde{y})\,
\sqrt{-\tilde{g}(\tilde{y})}\,-\,
\nonumber \\
&-&
\,\frac{m_{p}^{6}}{2}\,\int\, d^4 \tilde{x}\,\int\, d^4 \tilde{y}\,\sqrt{-g(\tilde{x})}\,f(\tilde{x})\,
\tilde{G}(\tilde{x},\tilde{y})\,f(\tilde{y})\,\sqrt{-g(\tilde{y})}\,-
\nonumber \\
&-&
m_{p}^{6}\,\xi\,\tilde{\xi}\int\, d^4 x\,\int\, d^4 \tilde{y}\,\sqrt{-g(x)}\,f(x)\,G(x,\tilde{y})\,
\tilde{f}(\tilde{y})\,\sqrt{-\tilde{g}(\tilde{y})}\,-
\nonumber \\
&-&
m_{p}^{6}\,\xi\,\tilde{\xi}\int\, d^4 x\,\int\, d^4 \tilde{y}\,\sqrt{-g(x)}\,f(x)\,\tilde{G}(x,\tilde{y})\,
\tilde{f}(\tilde{y})\,\sqrt{-\tilde{g}(\tilde{y})}\,,
\eeqar
for the simplicity we omit a dependence on the constants inside the r.h.s. of the expression.  
Now we see that there is an interaction term between the manifolds
in correspondence to \eq{CC3} definition. There are two cases we consider separately. 
For $\xi\,=-\,\tilde{\xi}\,=\,\pm\,1$ values of the parameters, we have:
\beqar\label{DS4}
\phi_{cl}\,& = &\,m_{p}^{3}\,G(x,y,\Lambda)\,\Le f(y)\sqrt{-g(y,\Lambda)}\,
 -\,\tilde{f}(y)\sqrt{-\tilde{g}(y,\tilde{\Lambda})}\,\Ra\,,
\\
\tilde{\phi}_{cl}\,& = &\,\pm\,m_{p}^{3}\,\tilde{G}(\tilde{x},\tilde{y},\tilde{\Lambda})\,
\Le \,f(\tilde{y})\sqrt{-g(\tilde{y},\Lambda)}\,
-\,\tilde{f}(\tilde{y})\sqrt{-\tilde{g}(\tilde{y},\tilde{\Lambda})}\,\Ra\,.
\eeqar
We see, that when 
$f(y)\sqrt{-g(y,\Lambda)}\,=\,\tilde{f}(y)\sqrt{-\tilde{g}(y,\tilde{\Lambda})}$ 
the expressions are trivial, but this equality is not requested in general and especially when 
the constants in $\sqrt{-g}$ are  non-zero and different, the \eq{Add1} symmetry is precise only for the zero cosmological constants.
Now, in order to define the cosmological constants from \eq{DS1} interaction term to the first perturbative order of precision, we will make the following. We 
will keep in \eq{DS3}  first order propagators, i.e. the flat ones, and will perform variation with respect to
$g$ and $\tilde{g}$ separately. Only after the variation will be taken we will set in the r.h.s. of the equations zero cosmological constants and will consider 
remaining integrals without their connection to $x$ and $\tilde{x}$ variables using \eq{Add1} symmetry. We will obtain then:
\beqar\label{DS5}
\Lambda(x)\,&=& \,-\,\frac{m_{p}^{4}}{4} \,f(x)\,\int\, d^4 y\, \sqrt{-g(y)}\,\Le G(x,y) + \tilde{G}(x,y) \Ra\,\Le \tilde{f}(y) -f(y) \Ra\,
\nonumber \\
\tilde{\Lambda}(x)\,& = &\,\frac{m_{p}^{4}}{4} \,\tilde{f}(x)\,\int\, d^4 y\, \sqrt{-g(y)}\,\Le G(x,y) +\tilde{G}(x,y)\Ra\,
\Le \tilde{f}(y) -f(y) \Ra\,.
\eeqar  
As we see, in flat manifold for the first kind of the scalar fields, \eq{Sec6} $\Lambda\,=\,\tilde{\Lambda}\,=\,0$ whereas for second type of the fields
we will obtain here $\Lambda\,=\,-\,\tilde{\Lambda}$ with relative sign depend on the sign of $\tilde{f} - f $ difference.

  Correspondingly, when $\xi\,=\,\tilde{\xi}\,=\,\pm\,1$, we obtain then:
\beqar\label{DS6}
\phi_{cl}\,& = &\,m_{p}^{3}\,G(x,y,\Lambda)\,\Le f(y)\sqrt{-g(y,\Lambda)}\,
 +\,\tilde{f}(y)\sqrt{-\tilde{g}(y,\tilde{\Lambda})}\,\Ra\,,
\\
\tilde{\phi}_{cl}\,& = &\,\pm\,m_{p}^{3}\,\tilde{G}(\tilde{x},\tilde{y},\tilde{\Lambda})\,
\Le \,f(\tilde{y})\sqrt{-g(\tilde{y},\Lambda)}\,
+\,\tilde{f}(\tilde{y})\sqrt{-\tilde{g}(\tilde{y},\tilde{\Lambda})}\,\Ra\,
\eeqar
and for the cosmological constants
\beqar\label{DS7}
\Lambda(x)\,&=& \,-\,\frac{m_{p}^{4}}{4} \,f(x)\,\int\, d^4 y\, \sqrt{-g(y)}\,\Le G(x,y) + \tilde{G}(x,y) \Ra\,\Le \tilde{f}(y) + f(y) \Ra\,
\nonumber \\
\tilde{\Lambda}(x)\,& = &\,-\,\frac{m_{p}^{4}}{4} \,\tilde{f}(x)\,\int\, d^4 y\, \sqrt{-g(y)}\,\Le G(x,y) +\tilde{G}(x,y)\Ra\,
\Le \tilde{f}(y) + f(y) \Ra\,.
\eeqar  
For the first kind of the scalar fields, \eq{Sec6}, for the flat case again $\Lambda\,=\,\tilde{\Lambda}\,=\,0$, whereas for second type of the fields
we  obtain  $\Lambda\,=\,\tilde{\Lambda}$ with the relative sign depend on sign of $f$.

\section{Possible forms of the interactions of the fields in the Lagrangian }
\label{S412}

 With the hekp of the  introduced CPTM symmetry we can unambiguously determine the possible forms of the interactions between the scalar fields in the Lagrangian. Namely, 
there is a request that the symmetry 
between the classical fields must be preserved in the general Lagrangian at the case of zero cosmological constants.
Therefore we can consider the connection of the possible interactions terms in the Lagrangian basing on the properties of classical fields and propagators of two 
flat manifolds.
Consider for example the first type of the scalar fields
\beq\label{CC71}
\tilde{\phi}_{cl}\,=\,-\phi_{cl}\,
\eeq
and connection between the propagators given by \eq{Sec6} which we suppose to be correct in the case of the flat manifold or in the case of the metric without the cosmological constant. Therefore,
the request of CPTM symmetry  leads to the following connection between the possible interaction terms in the Lagrangians of A and B manifolds:
\beq\label{CC9}
\left\{ 
\begin{array}{c}
\lambda_{\,2n+1}\,\phi^{\,2n+1}_{cl}(x)\,\rightarrow\,-\,\lambda_{\,2n+1}\,\tilde{\phi}^{\,2n+1}_{cl}(\tilde{x}) \\
\lambda_{\,2(n+1)}\,\phi^{\,2(n+1)}_{cl}(x)\,\rightarrow\,\lambda_{\,2(n+1)}\,\tilde{\phi}^{\,2(n+1)}_{cl}(\tilde{x}) 
\end{array}
\right.;\,\,\,
n\,=\,1,\,2,\,\cdots \,,
\eeq
in correspondence to the form of \eq{CC71} transformation.
Now we consider the second possibility of the connection between the classical fields:
\beq\label{CC10}
\tilde{\phi}_{cl}\,=\,
\phi_{cl}.
\eeq 
The consequence of that relation between the classical fields solutions is that now we have instead \eq{CC9}:
\beq\label{CC11}
\lambda_{\,n}\,\phi^{\,n}_{cl}(x)\,\rightarrow\,\lambda_{\,n}\,\tilde{\phi}^{\,n}_{cl}(\tilde{x})\,,\,\,\,n\,=\,3,\,4,\,\cdots\,.
\eeq
for the possible interacting terms of the fields in the Lagrangian of the approach.

  For the second choice of the scalar field in the problem, we have to 
put attention that in momentum space
\beq\label{CC14}
(-G_{D})\,=\,G_{F}^{\,*}\,,
\eeq
see Appendix A definitions.	
Therefore we have to the first approximation:
\beq\label{CC15}
\tilde{\phi}_{cl}\,=\,\phi_{cl}^{\,*}\,.
\eeq
The \eq{CC15} transformation rule, in turn, dictates the transformations of the possible interaction terms in the Lagrangians. We obtain:
\beq\label{CC161}
\imath\lambda_{\,n}\,\phi^{\,n}_{cl}(x)\,\rightarrow\,-\imath\lambda_{\,n}\,\tilde{\phi}^{\,n}_{cl}(\tilde{x})\,,\,\,\,n\,=\,3,\,4,\,\cdots\,,
\eeq
in full correspondence to the closed time path (Keldysh) formalism of non-equilibrium condensed matter physics, 
\cite{Shw,Kel} and \cite{Kel1} for example.
For the second case when
\beq\label{CC17}
\tilde{\phi}_{cl}\,=\,-\,\phi_{cl}^{\,*}\,
\eeq
we have to define
\beq\label{CC18}
\imath\lambda_{\,n}\,\phi^{\,n}_{cl}(x)\,\rightarrow\,\imath\,\Le\,-1\,\Ra^{n+1}\,\lambda_{\,n}\,\tilde{\phi}^{\,n}_{cl}(\tilde{x})\,,\,\,\,n\,=\,3,\,4,\,\cdots\,,
\eeq
in correspondence to \eq{CC17}.

For the further applications  we denote the Lagrangians considered here as cases $1-a,\, 1-b$ and $2-a,\, 2-b$ correspondingly to the order of their appearance in the Section. 

\section{Action of the formalism: interacting free fields}
\label{S42}

 Following the analogy with Keldysh formalism, now we introduce an interaction between the free fields of different manifolds. Again, itroducing in the Lagrangian corresponding interaction terms between the fields of different manifolds we consider separately the cases of two different types of scalar fields.
Also, here we do not include in the Lagrangian the source terms, they are not important for the further results. 

 First of all consider $1-a$ and $1-b$ Lagrangians in the case of flat A, B manifolds. As usual we define the following Wightman functions:
\beq\label{inf1}
\Delta_{>}(x,y)\,=\,-\imath\,D(x-y)\,,\,\,\,\tilde{\Delta}_{>}(\tilde{x},\tilde{y})\,=\,\imath\,D(\tilde{y}-\tilde{x})\,=\,-\,\Delta_{<}(\tilde{x},\tilde{y})\,,
\eeq
see \eq{vs9} definition. We, therefore, define the interaction part of the action  $S_{int}$ as:
\beq\label{inf2}
S_{int}\,=\,\frac{1}{2}\,
\int\, d^4 x\,\sqrt{-g}\,(\phi\,\,\tilde{\phi})\,
\left(
\begin{array}{c c}
G_{F}^{-1} & \Delta_{>}^{-1} \\
\tilde{\Delta}_{>}^{-1} & \tilde{G}_{F}^{-1}
\end{array}
\right)\,
\left(
\begin{array}{c}
\phi \\
\tilde{\phi}
\end{array}
\right)\,-\,
\int\, d^4 x\,\sqrt{-g}\,
(\phi\,\,\tilde{\phi})\,
\left(
\begin{array}{c}
 m_{p}^{3}  +J_{1}\\
\pm \Le m_{p}^{3}  +J_{2} \Ra
\end{array}
\right)\,.
\eeq
Here we introduced the auxiliary currents $J_{1}$ and $J_{2}$ in order to check the appearance of the additional Green's functions in the problem.
After the integration we obtain:
\beqar\label{inf4}
&\,& S_{int}\,= \,-\,\frac{m_{p}^{6}}{2}\,\int\, d^4 x\,\int\, d^4 y\,\sqrt{-g}\,G_{F}(x,y)\,\sqrt{-g}\,-\,
\frac{m_{p}^{6}}{2}\,\int\, d^4 \tilde{x}\,\int\, d^4 \tilde{y}\,\sqrt{-g}\,\tilde{G}_{F}(\tilde{x},\tilde{y})\,\sqrt{-g}\,-\,
\nonumber \\
&-&
\frac{m_{p}^{3}}{2}\int\, d^4 x \int\, d^4 y\,\sqrt{-g}\,J_{1}\,G_{F}(x,y)\,J_{1}\,\sqrt{-g}-
\frac{m_{p}^{3}}{2}\int\, d^4 \tilde{x} \int\, d^4 \tilde{y}\,\sqrt{-g} J_{2}\,\tilde{G}_{F}(\tilde{x},\tilde{y})\,J_{2}\sqrt{-g} -
\nonumber \\
&-&
\frac{m_{p}^{6}}{2}\int\, d^4 x \int\, d^4 y\,\sqrt{-g}  \Le \Delta_{>}(x,y) + \tilde{\Delta}_{>} (y,x) \Ra \sqrt{-g} \mp 
\nonumber \\
&\mp &
\frac{1}{2}\int\, d^4 x \int\, d^4 y\,\sqrt{-g} J_{2} \Le \Delta_{>}(x,y) + \tilde{\Delta}_{>} (y,x) \Ra J_{1} \sqrt{-g}  \mp
\nonumber \\
&\mp &
\frac{m_{p}^{3}}{2}\int\, d^4 x \int\, d^4 y\,\sqrt{-g}  \Le \Delta_{>}(x,y) + \tilde{\Delta}_{>} (y,x) \Ra J_{1} \sqrt{-g} \mp
\nonumber \\
&\mp &
\frac{m_{p}^{3}}{2}\int\, d^4 x \int\, d^4 y\,\sqrt{-g}  \Le \Delta_{>}(x,y) + \tilde{\Delta}_{>} (y,x) \Ra J_{2} \sqrt{-g}\,.
\eeqar
Taking into account that
\beq\label{inf5}
\Delta_{>}(x,y) + \tilde{\Delta}_{>} (y,x)\,=\,\Delta_{>}(x,y)\,-\,\Delta_{<}(y,x)\,=\,0\,,
\eeq
see\eq{inf1}, we obtain:
\beqar\label{inf6}
&\,& S_{int}\,= \,-\,\frac{m_{p}^{6}}{2}\,\int\, d^4 x\,\int\, d^4 y\,\sqrt{-g}\,G_{F}(x,y)\,\sqrt{-g}\,-\,
\frac{m_{p}^{6}}{2}\,\int\, d^4 \tilde{x}\,\int\, d^4 \tilde{y}\,\sqrt{-g}\,\tilde{G}_{F}(\tilde{x},\tilde{y})\,\sqrt{-g}\,-\,
\nonumber \\
&-&
\frac{m_{p}^{3}}{2}\int\, d^4 x \int\, d^4 y\,\sqrt{-g}\,J_{1}\,G_{F}(x,y)\,J_{1}\,\sqrt{-g}-
\frac{m_{p}^{3}}{2}\int\, d^4 \tilde{x} \int\, d^4 \tilde{y}\,\sqrt{-g} J_{2}\,\tilde{G}_{F}(\tilde{x},\tilde{y})\,J_{2}\sqrt{-g}\,.
\eeqar
The answer demonstrate that there is no Keldysh like interactions between the flat A and B manifolds in the case of $1-a$ and $1-b$ Lagrangians. If we will not introduce the interaction potential between
the scalar fields we will stay with two non-interacting scalar fields which provide the cancellation of mutual classical zero modes but not interact anymore. The situation can be different if we will account generated non-zero cosmological constants. In this case, thee quality \eq{inf5} can be not correct anymore, instead will have:
\beq\label{inf51}
\Delta_{>}(x,y,\Lambda) + \tilde{\Delta}_{>} (y,x,\tilde{\Lambda})\,=\,
( \D_{\Lambda}\Delta_{>}(x,y,\Delta) )_{\Lambda\,=\,0}\,\Lambda\,-\,
( \D_{\tilde{\Lambda}}\Delta_{<}(y,x,\tilde{\Lambda}) )_{\tilde{\Lambda}\,=\,0}\,\tilde{\Lambda}\,\neq\,0\,,
\eeq
and some interaction between the manifolds will arise.

 Now we consider $2-a$ and $2-b$ Lagrangians. Again we define the Wightman functions:
\beq\label{inf7}
\Delta_{>}(x,y)\,=\,-\imath\,D(x-y)\,,\,\,\,\tilde{\Delta}_{>}(\tilde{x},\tilde{y})\,=\,\imath\,D(\tilde{x}-\tilde{y})\,=\,-\,\Delta_{>}(\tilde{x},\tilde{y})\,,
\eeq
see \eq{vs9}. Repeating the same calculations as above we will obtain:
\beqar\label{inf8}
&\,& S_{int}\,= \,-\,\frac{m_{p}^{6}}{2}\,\int\, d^4 x\,\int\, d^4 y\,\sqrt{-g}\,G_{F}(x,y)\,\sqrt{-g}\,-\,
\frac{m_{p}^{6}}{2}\,\int\, d^4 \tilde{x}\,\int\, d^4 \tilde{y}\,\sqrt{-g}\,\tilde{G}_{F}(\tilde{x},\tilde{y})\,\sqrt{-g}\,-\,
\nonumber \\
&-&
\frac{m_{p}^{3}}{2}\int\, d^4 x \int\, d^4 y\,\sqrt{-g}\,J_{1}\,G_{F}(x,y)\,J_{1}\,\sqrt{-g}-
\frac{m_{p}^{3}}{2}\int\, d^4 \tilde{x} \int\, d^4 \tilde{y}\,\sqrt{-g} J_{2}\,\tilde{G}_{F}(\tilde{x},\tilde{y})\,J_{2}\sqrt{-g} -
\nonumber \\
&-&
\frac{m_{p}^{6}}{2}\int\, d^4 x \int\, d^4 y\,\sqrt{-g}  \Le \Delta_{>}(x,y) + \tilde{\Delta}_{>} (y,x) \Ra \sqrt{-g} \mp 
\nonumber \\
&\mp &
\frac{1}{2}\int\, d^4 x \int\, d^4 y\,\sqrt{-g} J_{2} \Le \Delta_{>}(x,y) + \tilde{\Delta}_{>} (y,x) \Ra J_{1} \sqrt{-g}  \mp
\nonumber \\
&\mp &
\frac{m_{p}^{3}}{2}\int\, d^4 x \int\, d^4 y\,\sqrt{-g}  \Le \Delta_{>}(x,y) + \tilde{\Delta}_{>} (y,x) \Ra J_{1} \sqrt{-g} \mp
\nonumber \\
&\mp &
\frac{m_{p}^{3}}{2}\int\, d^4 x \int\, d^4 y\,\sqrt{-g}  \Le \Delta_{>}(x,y) + \tilde{\Delta}_{>} (y,x) \Ra J_{2} \sqrt{-g}\,.
\eeqar
Due the \eq{inf7} identity now we have:
\beq\label{inf9}
\Delta(x,y)\,=\,\Delta_{>}(x,y) + \tilde{\Delta}_{>} (y,x)\,=\,\Delta_{>}(x,y)\,-\,\Delta_{>}(y,x)\,=\,\Delta_{>}(x,y)\,-\,\Delta_{<}(x,y)\,\neq\,0\,
\eeq
already in case of zero cosmological constant.
Therefore, we see, in full analogy with Keldysh formalism, that there is a new term of direct interaction between the A and B manifolds arises at the case of zero auxiliary currents:
\beqar\label{inf10}
&\,& S_{int}\,= \,-\,\frac{m_{p}^{6}}{2}\,\int\, d^4 x\,\int\, d^4 y\,\sqrt{-g}\,G_{F}(x,y)\,\sqrt{-g}\,-\,
\frac{m_{p}^{6}}{2}\,\int\, d^4 \tilde{x}\,\int\, d^4 \tilde{y}\,\sqrt{-g}\,\tilde{G}_{F}(\tilde{x},\tilde{y})\,\sqrt{-g}\,-\,
\nonumber \\
&-&
\frac{m_{p}^{6}}{2}\int\, d^4 x \int\, d^4 \tilde{y}\,\sqrt{-g(x)}  \,\Delta(x,\tilde{y}) \, \sqrt{-g(\tilde{y})}\,.
\eeqar
This interaction between the A and B  manifolds takes place at quantum level with non-zero Keldysh propagator appearance in the calculations.

\section{Action of the formalism: self-interacting fields}
\label{S5}

  Our next step is an introduction of self-interactions of the fields. The connection between the corresponding parts of the two Lagrangians are given by 
\eq{CC9}, \eq{CC11} and \eq{CC161}, \eq{CC18}. Correspondingly to that we will obtain quantum corrections to the classical values of the fields given by \eq{CC5}.
As a result of integration around the classical solutions \eq{CC5} we will obtain an effective action of the following general form:
\beqar\label{Self1}
\Gamma_{int} & = &\sum_{m,n=1}^{\infty}\int d^{4} x_{1}\sqrt{-g}\cdots\int d^{4} x_{n}\sqrt{-g}
\int d^{4} \tilde{x}_{1}\sqrt{-\tilde{g}}\cdots\int d^{4} \tilde{x}_{m}\sqrt{-\tilde{g}}\Le
V_{0}(x_{1},\cdots ,x_{n};\tilde{x}_{1},\cdots,\tilde{x}_{m})+
\right. \nonumber  \\
&+&
\left.
V_{1\,\cdots\,n;1\,\cdots\,m}(x_{1},\cdots,x_{n};\tilde{x}_{1},\cdots,\tilde{x}_{m})\,
\phi_{cl\,1}\,\cdots\,\phi_{cl\,n}\,\tilde{\phi}_{cl\,1}\,\cdots\,\tilde{\phi}_{cl\,m}\,\Ra\,.
\eeqar
In this case the cosmological constant can be defined mostly general as
\beqar\label{Self2}
\Lambda(x) & = &\frac{1}{m_{p}^{2}}
\sum_{m,n=1}^{\infty}\int d^{4} x_{1}\sqrt{-g}\cdots\int d^{4} x_{n}\sqrt{-g}
\int d^{4} \tilde{x}_{1}\sqrt{-\tilde{g}}\cdots\int d^{4} \tilde{x}_{m}\sqrt{-\tilde{g}}
\Le V_{0}\,+\,
\right. \nonumber  \\
&+&
\left.
V_{1\,\cdots\,n;1\,\cdots\,m}(x,x_{1},\cdots,x_{n};\tilde{x}_{1},\cdots,\tilde{x}_{m})\,
\phi_{cl\,1}\,\cdots\,\phi_{cl\,n}\,\tilde{\phi}_{cl\,1}\,\cdots\,\tilde{\phi}_{cl\,m}\,\Ra\,.
\eeqar
with $V_{0}$ as vacuum contributions to the constant (no-legs diagrams). Similarly to the previous expressions, the r.h.s. of \eq{Self1} and \eq{Self2}
also depend on the constants through the vertices and $\sqrt{-g}$, the \eq{Self2} is a non-linear equation for the constant.

  In the effective action, beginning from $n\,=\,2$ external legs, the introduced vertices will contribute to the renormalization of the theory determining 
renormalized mass and vertices of the scalar field. 	
The mostly unpleasant
contributions there are the vacuum ones and quantum contributions to the $\phi_{cl}$ (diagrams with one external leg present\footnote{Further we will consider $\phi^{4}$ theory without the quantum contribution to $\phi_{cl}$.}).
As we mentioned in the Introduction, the contributions to the cosmological constant in the formalism can be divided on the two parts. The first type of the 
contributions is due the quantum vacuum zero modes,
i.e. related to $V_{0}$ or quantum corrections to $\phi_{cl}$, the separated contributions are due the non-zero curvature of the A, B manifolds. There are also quantum contributions in non-flat manifolds, we do not consider these effects.  Namely, further in the Section,  we will neglect the curvature corrections to the propagators in the calculations of \eq{Self1} one-loop effective action and will consider the theory in the flat space-time.

\subsection{Self-interacting scalar fields of the first kind}

 We again begin from $1-a,b$ Lagrangians adding to \eq{CC6} the self-interaction terms accordingly to \eq{CC9} prescription. We will consider only renormalizable 
theories, for the simplicity limiting ourselves by the consideration of $\phi^{4}$ power of interaction in $D=4$ dimensions and by the one-source Lagrangian.  We will obtain for the action:
\beqar\label{Self11}
S_{int}\,& = &\,\int\, d^4 x\,\sqrt{-g}\,\Le \,\frac{1}{2}\,\phi\,G_{F}^{-1} \,\phi \,-\,m_{p}^{3}\,\phi\,\Ra\,+\,
\int\, d^4 \tilde{x}\,\sqrt{-g}\,\Le\,\frac{1}{2}\, \tilde{\phi}\,\tilde{G}_{F}^{-1} \,\tilde{x}\,\tilde{\phi} \,\mp\,m_{p}^{3}\,\tilde{\phi}\,\Ra\,-\,
\nonumber \\
&-&\,
\int\, d^4 x\,\sqrt{-g}\,\lambda_{4}\,\frac{\phi^{4}(x)}{4\,!}\,-\,\int\, d^4 \tilde{x}\,\sqrt{-g}\,\lambda_{4}\,\frac{\tilde{\phi}^{4}(\tilde{x})}{4\,!}\,.
\eeqar
Now we expand the potential of the self-interaction around the classical values of the fields and preserving in the expressions terms till quadratic with respect to fluctuations  obtain:
\beqar\label{Self3}
S_{int}\,& = &\,-\,
\int\, d^4 x\,\lambda_{4}\,\frac{\phi^{4}_{cl}(x)}{4\,!}\,-\,\int\, d^4 \tilde{x}\,\lambda_{4}\,\frac{\tilde{\phi}^{4}_{cl}(\tilde{x})}{4\,!}\,+\,
\nonumber \\
&+&\,
\frac{1}{2}\,\int\, d^4 x\,\varepsilon\,G_{F}^{-1} \,\varepsilon \,-\,\frac{m_{p}^{6}}{2}\,\int\, d^4 x\,\int\, d^4 y\,\sqrt{-g(x)}\,G_{F}(x,y)\,\sqrt{-g(y)}\,+
\nonumber \\
&+&
\,\frac{1}{2}\,\int\, d^4 \tilde{x}\,\tilde{\varepsilon}\,\tilde{G}_{F}^{-1} \,\tilde{\varepsilon} \,-\,
\frac{m_{p}^{6}}{2}\,\int\, d^4 \tilde{x}\,\int\, d^4 \tilde{y}\,\sqrt{-g(\tilde{x})}\,\tilde{G}_{F}(\tilde{x},\tilde{y})\,\sqrt{-g(\tilde{y})}\,-\,
\nonumber \\
&-&
\int\, d^4 x\,\Le\, \lambda_{4}\,\frac{\phi^{3}_{cl}\,\varepsilon}{3\,!}\,+\,\lambda_{4}\,\frac{\phi^{2}_{cl}\,\varepsilon^2}{4}\,\Ra\,-\,
\int\, d^4 \tilde{x}\,\Le \, \lambda_{4}\,\frac{\tilde{\phi}^{3}_{cl}(\tilde{x})\,\tilde{\varepsilon}}{3\,!}\,+\,
\lambda_{4}\,\frac{\tilde{\phi}^{2}_{cl}(\tilde{x})\,\tilde{\varepsilon}^{2}}{4}\,\Ra\,.
\eeqar
Without the contribution of classical fields and cosmological constant terms, we obtain after the integration with respect to the fluctuations:
\beqar\label{Self4}
S_{int}= &-&\frac{\lambda_{4}^{2}}{2\,(3\,!)^{2}}\,\int d^4 x\,\int d^4 y \, (\phi^{3}_{cl}(x))\,G_{F}(x,y)\,(\phi^{3}_{cl}(y))\,-
\nonumber \\
&-&\,
\frac{\lambda_{4}^{2}}{2\,(3\,!)^{2}}\,\int  d^4 \tilde{x}\,\int  d^4 \, \tilde{y}\, (\tilde{\phi}^{3}_{cl}(\tilde{x}))\,\tilde{G}_{F}(\tilde{x},\tilde{y})\,
(\tilde{\phi}^{3}_{cl}(\tilde{y}))\, +
\nonumber \\
&+&\,
\frac{\imath}{2}\,Tr\,\ln \Le 1\,+\,\frac{\lambda_{4}}{4}\,G_{F}\,\phi^{2}_{cl}\,\Ra\,+\,
\frac{\imath}{2}\,Tr\,\ln \Le 1\,+\,\frac{\lambda_{4}}{4}\,\tilde{G}_{F}\,\tilde{\phi}^{2}_{cl}\,\Ra\,.
\eeqar
Due the \eq{Sec5} relation between the Green's function, the two first terms in \eq{Self4} cancel each 
another\footnote{In the case of the non-flat manifold, when we distinguish between $g$ and $\tilde{g}$ metrics, these two terms will provide corrections to the value of cosmological constant in 
 A, B manifolds. } 
and we remain with the following expression to one-loop order precision:
\beq\label{Self5}
S_{int}= 
\frac{\imath}{2}\,Tr\,\ln \Le 1\,+\,\frac{\lambda_{4}}{4}\,G_{F}\,\phi^{2}_{cl}\,\Ra\,+\,
\frac{\imath}{2}\,Tr\,\ln \Le 1\,+\,\frac{\lambda_{4}}{4}\,\tilde{G}_{F}\,\tilde{\phi}^{2}_{cl}\,\Ra\,.
\eeq
We see, therefore, that the relative sign of the classical solution does not matter in the case of flat manifolds, we can take further $\phi_{cl}\,=\,\tilde{\phi}_{cl}$.
The only difference in the contributions of the terms are due the different sign of Green's functions that affect only on diagrams with odd number of propagators which will be canceled in the answer.
The same rules are applicable in the case of many-loops diagrams. We are interesting mostly in the two-loops vacuum diagram without external legs provided by
$\lambda_{4}\,\varepsilon^{4}$ terms in the potential.
This diagram is not zero without the regularization and doubled in the final answer. 

 The situation is more interesting if instead potential with separate self-interacting fields we will consider  a potential similar to the $\phi^{4}$ potential of the scalar doublet:
\beq\label{Self6}
V(\phi,\,\tilde{\phi})\,=\,\frac{\lambda_{4}}{4\,!}\,\Le\,\phi^{2}\,+\,\tilde{\phi}^{2}\,\Ra^{2}\,=\,V_{A A}(\phi)+V_{A B}(\phi,\tilde{\phi})+V_{B B}(\tilde{\phi})\,.
\eeq
The consequence of the appearance of the $\frac{\lambda_{4}}{3}\,\phi_{cl}\,\tilde{\phi}_{cl}\,\varepsilon\,\tilde{\varepsilon}$ quadratic term in the potential is that
there is now a mixing of the propagators of the problem
\beq\label{Self7}
G_{\mu \nu}\,=\,G_{F\,\mu \nu}\,-\,G_{F\,\mu \rho}\,V_{\rho \sigma}^{''}(\phi\,\tilde{\phi})\,G_{\sigma \nu}\,,\,\,\,\mu\,=\,A,B\,,
\eeq
with $V_{\rho \sigma}^{''}(\phi\,\tilde{\phi})$ as coefficients of the quadratic with respect to $\varepsilon$ and $\tilde{\varepsilon}$ terms.
The answer for the one-loop action term in this case we can write as
\beq\label{Self8}
S_{int}= 
\frac{\imath}{2}\,Tr\,\ln \Le 1\,+\,\frac{\lambda_{4}}{3}\,G_{F\,\mu \nu}\,\phi_{\nu\, cl}\,\phi_{\rho\, cl}\Ra\,.
\eeq
This mixing of the propagators in the diagrams is arising beginning from the diagrams of the $\lambda_{4}^{2}$ and higher orders, independently on the number of loops. As a result there is no 
quantum corrections in the problem, in this formulation of the formalism the quantum corrections in two flat manifolds are odd with respect to 
the CPTM symmetry transform giving an overall zero in the final answer. The mostly interesting consequence of that for us it is zero contribution of the vacuum diagrams into the cosmological constant in the flat space-time.
Namely, it can be clarified as following.  For these kind of the diagrams at $\lambda_{4}^{N},\,N\,=\,1,\,\cdots$ order there are $N$ pairs of  propagators over all. Each pair of propagators in turn
gives  zero contribution in the diagram because there are
two positive and two negative expressions for each pair of propagators.

\subsection{Self-interacting scalar fields of the second kind}

 We will not consider here the well-known general closed time path formalism for the scalar fields in the flat manifolds, the different applications of the framework can be found in \cite{Kel1} for example. Instead we will discuss only two-loop no-legs vacuum diagram which contribute into the cosmological constant value in the case of interacting doublet fields.
We have for the potential of the problem:
\beq\label{Self9}
V(\phi,\,\tilde{\phi})\,=\,\frac{\lambda_{4}}{4\,!}\,\Le\,\phi^{2}\,-\,\tilde{\phi}^{2}\,\Ra^{2}\,=\,V_{A A}(\phi)-V_{A B}(\phi,\tilde{\phi})+V_{B B}(\tilde{\phi})\,,
\eeq
see \eq{CC161} and \eq{CC18} identities.
There are two separated contributions we can write on the base of corresponding Feynman rules:
\beqar\label{Self10}
V_{0,1,2-loops}\,& \propto &\,(G_{F}(x,x)\,-\,\tilde{G}_{F}(x,x))^2\,=\,(G_{F}(x,x)\,+\,G_{D}(x,x))^2\,\propto\,
\nonumber \\
&\propto\,&\frac{1}{p^2 - m^2-\imath\varepsilon}\,-\,\frac{1}{p^2 - m^2+\imath\varepsilon}\,=\,
\frac{2\imath\varepsilon}{(p^2 - m^2)^2 + \varepsilon^2}\,\stackrel{\varepsilon\rightarrow 0}{\longrightarrow}\,0
\eeqar
and
\beq\label{Self1101}
V_{0,2,2-loops}\, \propto \,\Delta^{2}(x,x)\,=\,(\Delta_{>}(x,x)\,-\,\Delta_{<}(x,x))^2\,=\,0\,
\eeq
see \eq{inf9} definition. We will not discuss here the diagrams of $\lambda_{4}^{2}$ or higher orders where mix of the $G$ and $\Delta$ propagators
will happen as well, 
see \cite{Kel1} and references therein for example. So we see that the contribution of the vacuum diagrams to the cosmological constant in the flat manifolds is zero up to the two loops precision at least.

\section{Smallness of the cosmological constant}
\label{S6}

 Now we consider, first of all, the possible quantum contributions to the cosmological constant value. In the formalism the constant is evolving with time,
therefore we will discuss the value of the constant in an almost flat space-time and will not consider the possibilities of the large value of the constant for the
manifolds with large curvature. It was demonstrated in the previous Sections that the vacuum contributions into the cosmological constant value are 
depend on the form of the scalar fields and interactions between them introduced. In the formalism, nevertheless, there is a possibility to eliminate all types of the vacuum diagrams in general. Let's consider the partition function for the scalar field in the curved space-time defining it as following: 
\beq\label{Sml1}
Z[f]=Z^{-1}[f=0]\,\int D\phi\,D \tilde{\phi}\,e^{\imath\,S_{int}(\phi, \tilde{\phi}, f)}\,,
\eeq
with $S_{int}$ given by \eq{CC4} for example or by any another action with interaction between the fields introduced. In this definition we used the fact that
the currents $f$ are physical ones and we can define the normalization factor for the fields without the currents also in the curved manifold. Clearly, this definition eliminates the all vacuum diagrams from the consideration without the relation to the curvature of the manifolds. 

 Therefore we stay with the smallness of the bare value of the constants given by \eq{CC801} or \eq{CC8031} expressions for the case of the
one-source Lagrangian or by \eq{DS5} and \eq{DS7} in the case of two-source Lagrangian. We assume, that the most natural scenario we can propose is that
the bare value of the cosmological constant is zero for the cases of flat A and B manifolds. In this case the simplest way to provide the result is to consider the massless scalar fields
with the $f\,=\,\tilde{f}\,=\,1$ sources. The proper regularization of the corresponding propagators gives
\beq\label{New1}
\int\,d^4\,y\,G_{0}(x,y)\,=\,0\,
\eeq
result for any  bare propagator of the fields. The non-zero value for the cosmological constant in this set-up will be exclusively due curvature corrections to the propagators
and some quantum corrections in the form of the effective vertices, all of them will be small in the almost flat space-time. Still, for the case of 
\eq{DS5} and \eq{DS7} the zero bare constants can be provided by the opposite values of the $f$ and $\tilde{f}$ sources. 

  In the case of massive field, without special correspondence between the $f$ and $\tilde{f}$ sources,
we again can begin the consideration from the flat space-time 
and $f=1$. 
We will obtain then:
\beq\label{Sml2}
\Lambda\,=\,\frac{m_{p}^{4}}{4\,m^2}\,
\eeq
that means 
\beq\label{Sml3}
m^2\,>>\,m_{p}^{2}\,,
\eeq
the sign of the constant here corresponds to it's final sign in the l.h.s. of the Einstein equations with Einstein-Hilbert action given by \eq{CC2}:
\beq\label{Sml31}
G_{\mu \nu}\,+\,g_{\mu \nu}\,\Lambda\,=\,0\,.
\eeq
We see that for the one-source Lagrangian the constant has wrong sign.
It must be noted, nevertheless, that in the case of the first type of scalar field, the bare values of the constants cancels each another in the action. The contribution in the equation of motion will come only from the second order contribution to the constants, which is squared with respect to the Appendix C integrals.
Therefore, in this case, the relative sign of the cosmological constant will be correct.
Nevertheless, in order to reproduce the known absolute value of the constant, we can define the mass of the scalar fields  in terms of the $m_{p}^{2}$
and some characteristic length $\lambda$:
\beq\label{Sml4}
m^2\,=\,m_{p}^{2}\,e^{\,\lambda^2\,m_{p}^{2}}\,,\,\,\,\lambda^2\,=\,\frac{1}{m_{p}^{2}}\,\ln(\frac{m_{p}^{2}}{4\,\Lambda})\,.
\eeq
we see that the mass is super-heavy at least at the present of the space-time.

 An another possibility to provide the smallness of the bare constants for the massive fields is through a smallness of the sources of the particles. Taking the mass of the
fields $m=m_{p}$ for simplicity, we assume:
\beq\label{Sml32}
f\,\propto\,\frac{R}{m^{2}_{p}}
\eeq
with $R$ as curvature of the manifold. In this  case we obtain:
\beq\label{Sml33}
\Lambda\,\propto\,\frac{R^2}{m^{2}_{p}}
\eeq
and this value is small simply because the flatness of the manifold.
The correct sign of the cosmological constant we obtain in the case of the interaction Lagrangian with two types of the sources for the same field.
First of all, consider the case of positive $f$ in  \eq{DS1}. Requiring 
\beq\label{Sml34}
\tilde{f}\,-\,f\,\propto\,\tilde{R}\,-\,R\,=\,\delta\,>\,0\,
\eeq
in \eq{DS5} and taking $\delta$ without loss of generality as some constant, we will obtain for the second type of the scalar fields:
\beq\label{Sml35}
\Lambda(x)\,=\,\frac{R\,\delta\,}{2\,m^{2}_{p}}\,
\eeq
with correct Einstein equations:
\beq\label{Sml36}
G_{\mu \nu}\,-\,g_{\mu \nu}\,\Lambda\,=\,0\,.
\eeq
The same result we can achieve when the source $f$ is negative. In this case we take \eq{DS6}-\eq{DS7} solution and assume that
\beq\label{Sml361}
\tilde{f}\,+\,f\,\propto\,\delta\,<\,0\,
\eeq
that again provide the correct sign of the constant.
These two possible mechanisms of the constant's smallness can be applied in this case as well. At the case of the first field, the bare value
of the constants is precisely zero and some corrections to it must be calculated basing on the expansion of the propagators and metric's determinant with respect to the curvature. 

  Concerning the possible origin of the field with the large mass, we can speculate about the following. 
Let's suppose that there are quantum fluctuations of the scalar curvature which are not accounted by the classical Einstein equations:
\beq\label{Sml5}
R\,\rightarrow\,R\,+\,\delta\,R\,=\,R\,+\,\,m_{p}\,\phi\,f\,=\,R\,+\,\phi\,\frac{R}{m_p}\,.
\eeq
These fluctuations indeed reproduce term similar to the source terms in \eq{CC4} and \eq{Sml32}. After that, considering the fluctuation as a regular scalar field, we can
 write Lagrangian \eq{CC4} for the field with one understandable restriction. The field can not propagate far and that is what required for the extremely heavy mass in the Lagrangian.

\section{Some conjectures about possible solution of singularity problem}
\label{S8}

 In general relativity there are a few possible types of singularities arise at classical and quantum levels. Perhaps, the full solution of the problem is impossible without the quantum gravity theory, which so far is not known. Therefore we will operate with classical gravity theory, there is nothing else we have, assuming that some main ideas of the approach will remain when the full gravity theory will be constructed. Further we present some oversimplified consideration of the consequences of the approach to the problems of black hole's and Big Bang singularities.

 The first singularity we discuss it is the singularity of the metric of isolated Schwarzschild black hole. We can assume, that
the mass parameter, which arises in the solution at vicinity of $r\,\rightarrow\,0$ singularity, is a function of the radial coordinate
\footnote{We use here $m$ as notation for a mass 
of some infinitely small volume around a singularity
in order to distinguish it from the $M$ parameter in the metric's solution far away from the gravitating mass.}:
\beq\label{Sin1}
m\,=\,m(r)\,=\,m_{0}\,+\,m_{1}\,r\,+\,m_{2}\,r^2\,+\,m_{3}\,r^3\,+\,...\,,\,\,\,r\,\rightarrow\,0\,
\eeq
with $m_{i}$ as some coefficients of the expansion,
see \cite{Nov, Skok, Frol} for the similar examples.
Now, we require that the mass will continuously change the sign under the $r\,\rightarrow\,-r$ transform and therefore 
will require
instead \eq{Sin1}:
\beq\label{Sin2}
m\,=\,m(r)\,=\,m_{1}\,r\,+\,m_{3}\,r^3\,+\,...\,=\,\sum_{n=1}^{\infty}\,m_{2n+1}\,r^{2n+1}\,.
\eeq
Considering the Schwarzschild's metric we will obtain correspondingly:
\beq\label{Sin3}
g_{00}\,=\,f(r)\,=\,(1\,-\,2\,m_{1})\,\Le 1\,-\,2\,m_{3}\,\frac{r^{2}}{1\,-\,2\,m_{1}}\,+\,\cdots\,\Ra\,.
\eeq
Our next, natural, assumption is that $1\,-\,2\,m_{1}\,\neq\,0$ that allows to perform the variables change:
\beq\label{Sin4}
t\,\sqrt{1\,-\,2\,m_{1}}\,\rightarrow\,t\,,\,\,\,\,r\,/\,\sqrt{1\,-\,2\,m_{1}}\,\rightarrow\,r\,
\eeq
that will provide for the metric:
\beqar\label{Sin5}
ds^2\,& = &\,(1 - 2 m_3\,r^2\,+\cdots\,)\,dt^2\,-\,(1 - 2 m_3\,r^2\,+\cdots\,)^{-1}\,dr^2\,-\,(1 - 2 m_1)\,r^2\,
\Le d \theta^2\,+\,\sin^2 \theta\,d \phi^2\,\Ra\,=\, \\
&=&\,
(1 - 2 m_3\,r^2\,+\cdots\,)\,dt^2\,-\,(1 - 2 m_3\,r^2\,+\cdots\,)^{-1}\,dr^2\,-\,r^2\,d \Omega^{'2}\,.
\eeqar
The obtained metric, to leading order in expansion with respect to the powers of $r$, is similar to the 
static local parametrization of the de Sitter space.
Correspondingly, taking into account that the change of the value of the 2-d volume does not change 
the spherical symmetry of te metric, we will obtain for the scalar curvature at vicinity of $r\,=\,0$ the well known answer for the de Sitter space:
\beq\label{Sin6}
R\,=\,-24\,m_{3}\,=\,-12\,/\,R^{2}\,,
\eeq
with curvature
\beq\label{Sin7}
k\,=\,\frac{1}{R^2}\,=\,2\,m_{3}
\eeq
which can be positive or negative depending on the sign of $m_3$. We see, that the request of the analyticity of the metric at 
$r\,\rightarrow\,0$ is satisfied if we will assume the continuous change of the mass sign at $r\,\rightarrow\,-r$ transition 
through $r\,=\,0$ singularity.

 An another singularity we shortly discuss in the given framework is the Big Bang singularity. We make the following propositions
which are natural in the model. First of all, we consider the second Friedmann equation at vicinity $t\,=\,0$ point only and do not 
try to obtain any analytical solution valid at later time.
Second of all, we assume that the Universe is vacuum dominated at the initial moments of time's evolution, all other possible contributions 
in the equation arise at later moments of time and are not relevant for us.
The third proposition
is the consequence of the formalism: we do not separate between the cosmological constant and vacuum energy density in the equation
taking the  function in the r.h.s. of the equation as an overall vacuum energy density of our part of the manifold, which as well is an initial cosmological constant in the approach. This density is negative and depends on time, the proposed sign of the function is dictated by the anti-gravity repulsion between the parts of the extended manifold, the dependence on time
must be introduced because of the observation that the interaction between the parts of the general manifold should decrease during the evolution and because of the general structure of the constant in the proposed framework. Consequently, we will obtain for the second Friedmann equation for the only our part of the manifold the following expression:
\beq\label{Sin8}
\frac{\ddot{a}}{a}\,=\,\Lambda(t)\,,
\eeq
for simplicity we absorbed all the coefficients in the definition of $\Lambda$.
Now, let's consider a couple of simple models for $\Lambda(t)$ function and corresponding results for the scaling factor $a(t)$ in FRW metric.

 The first example we take is the following one:
\beq\label{Sin9}
\Lambda\,=\,\Lambda_{0}\,/\,t^{\beta}\,.
\eeq
Here we perform the following change of the variable:
\beq\label{Sin10}
z\,=\,t^{\alpha}\,,\,\,\,a\,=\,a(z)\,
\eeq
that provides for \eq{Sin8}
\beq\label{Sin11}
\frac{\alpha\,(\alpha -1)}{t^{2-\alpha}}\,\frac{d a}{d z}\,+\,\frac{\alpha^2}{t^{2-2\alpha}}\,\frac{d^2 a}{d z^2}\,=\,
a(z)\,\frac{\Lambda_{0}}{t^{\beta}}\,.
\eeq
We look for an analytical solution at vicinity of $t\,=\,0$, therefore we have:
\beqar\label{Sin12}
&\,&\,
2\,-\,\alpha\,>\,2\,-\,2\,\alpha\,\,\,\,\,\rightarrow\,\,\,\,\alpha\,>\,0 \\
&\,&\,2\,-\,\alpha\,=\,\beta\,\,\,\,\rightarrow\,\,\,\,\alpha\,=\,2\,-\,\beta\,>\,0\,\,\,\,\,\rightarrow\,\beta\,<\,2\,.
\eeqar
Correspondingly, for the $\beta\,<\,2$ we obtain at $t\,\rightarrow\,0$:
\beq\label{Sin121}
\alpha\,(\alpha -1)\,\frac{d a}{d z}\,=\,a(z)\,\Lambda_{0}\,
\eeq
that in turn provides
\beq\label{Sin13}
a(t)\,=\,a_{0}\,e^{\,\frac{\Lambda_{0}\,t^{2\,-\,\beta}}{ (2\,-\,\beta)\,(1\,-\,\beta)}}\,.
\eeq
We see, that there is an analytical inflation scenario in the example when
\beq\label{Sin14}
\Lambda_{0}\,>\,0\,,\,\,\,\,1\,-\,\beta\,>\,0
\eeq
or when
\beq\label{Sin15}
\Lambda_{0}\,<\,0\,,\,\,\,\,1\,-\,\beta\,<\,0\,.
\eeq
The another example we can discuss is for the following form of the $\Lambda$ function:
\beq\label{Sin16}
\Lambda\,=\,\Lambda_{0}\,e^{-\,\beta\,t^{\gamma}}\,\approx\,\Lambda_{0}\,\Le 1 -\beta\,t^{\gamma} \Ra\,.
\eeq
We again introduce new variable:
\beq\label{Sin17}
z\,=\,t^{2}\,+\,f(t)\,,\,\,\,f(t)\,\propto\,t^{\alpha}\,,\,\,\,\alpha\,>\,2\,.
\eeq
Therefore, at vicinity of $t\,\rightarrow\,0$ we can write 
\beq\label{Sin18}
2\,\frac{d a}{ d z}\,=\,a(z)\,\Lambda_{0}
\eeq
that provides a solution:
\beq\label{Sin19}
a(t)\,=\,a_{0}\,e^{\frac{\Lambda_{0}}{2}\,t^2}
\eeq
which is as well an analytic inflation solution for the scale factor at the case of $\Lambda_{0}\,>\,0$.

 We notice, that in general we have to consider the problem of singularity simultaneously for the both parts of the extended manifold. There are 
different possible scenario arise in this case, which depend on the directions of the time's arrows and initial values of the cosmological constants for the separated manifolds. We postpone this task for the separate publication.

 There are also additional types of singularities present in the general relativity. For example, there are soft singularities, 
see \cite{Kamen}, which can be crossed smoothly when some additional properties of the matter are assumed.
We note, see \cite{Chapl} for example, that in this models quite naturally arise a notion of the negative energy density 
of the matter, that, in some extend, is similar to the proposed in the paper. In general, therefore, it will be interesting to understand the intersections of the approaches which operate with ideas of some extended 
cosmology picture.

\section{Summary and discussion}
\label{S7}

  Let's us summarize the main propositions and results of the formalism. First of all, in the approach we introduced and considered the different manifolds A and B 
related by the CPTM symmetry transform. These two, or more, manifolds are parts of the extended solution of
the classical Einstein's equations, the interpretation of the second manifold as populated by the negative gravitational mass   
is an usual one in fact, 
the discussion of this issue, for example,  can be found in \cite{Chandr}. The proposed model also has some similarities to 
CPT symmetric Universe model considered in \cite{CPT} and models of \cite{Sym}. There are as well 
two-time direction models proposed for the solution of the Universe's low initial entropy value, see \cite{Saharov} and
more sophisticated two-times direction models of \cite{Bar}, which structure, nevertheless, is quite different from proposed here.
The idea of negative energy/anti-gravity regions in some type of an extended solution  of Einstein equiation
also have been considered in the literature, see for example \cite{Bars1,AntiG}. In general, of course, there are some similarities
and differences of these approaches with the formalism proposed here, it will be interesting to analyze the intersection points of the formalisms.

  The main consequences  of the set-up proposed in the paper is that we always have a system of two fields related by the symmetry and with different time directions in each manifold.
We also note, that unlike to \cite{Linde}, for example, the different terms in the general action are present with the same sign, see \eq{Dop1}-\eq{Dop2}.
The difference between the terms in the action is
appearing due the CPTM symmetry request applied to the scalar fields, for the metric the symmetry is precise when the cosmological constant is zero.

 The next basing part of the formalism is an additional bi-scalar term in the general action which "glue" the manifolds, see \eq{CC1} and \eq{CC3}. This 
term provides an effective interaction between the points of extended manifold, the cosmological constant arises there as a consequence of these interactions. In general, we can not define a priori the form of the interaction term, the simplest and mostly obvious way to introduce this term is to define it through a propagator of scalar field
which connects different points of the extended general solution, kind a quantum wormhole in some sense.

 The choose of the propagator of the scalar field as the interaction term  is not arbitrary of course. The introduced CPTM symmetry transforms the usual scalar field
of the A manifold into an another scalar field of B manifold. The properties of the quantized B-field are different from the A-field, there are also two possibility to determine the B-field.
Each possibility depends on the type of the closed time path defined for A-B manifolds together in correspondence to the positions of $in$ and $out$ states for the A,B fields. 
Nevertheless, in spite to the different propagators of two B fields, the main consequence of the application of the CPTM transform is that the overall 
"classical" zero modes contribution of A and B fields together in the energy-stress tensor is precise zero. This is a first important consequence of the proposed symmetry, it solves the zero mode problem of the cosmological constant.

 Thereby, the action for the two fields is based on the closed time paths. There are two variants of the time paths considered in the paper which correspond to the two types of the B scalar field. The second variant we considered is the same as the time path in the Keldysh - Schwinger, \cite{Shw,Kel}  $in$-$in$  QFT formalism, see 
also \cite{Kel1}. The next step we made is a construction of the QFT for the A and B fields in A and B manifolds. We again considered only two obvious possibilities for the corresponding QFT, both based on the choose of Feynman propagator $G_{F}$ as propagator of the scalar field of A manifold and and two different propagators for the B field obtained
after the application of the CPTM transform to $G_{F}$. In both cases the full
QFT can be constructed, with the self-interaction vertices of the fields and direct interaction between the A,B manifolds included. In the case of the direct
interaction between the A, B manifolds, the Keldysh propagator is revealed in the formalism.
It is demonstrated, that due the presence of the physical sources in the Lagrangian, in the partition function we can eliminate the quantum vacuum contributions into the cosmological constants values
ending with only contributions from the vertices of the quantum effective action of corresponding QFT, 
see \eq{Self1}-\eq{Self2}. In both cases the leading perturbative contribution to the constant is provided by the first term of the \eq{Self2} effective action.

 We obtain, therefore, that in the case of the simplest Lagrangian with one source for the fields, \eq{CC4}, the constant has a wrong sign
for the A manifold for the two types
the B scalar field. In turn, in the case of the Lagrangian with two sources for each field, \eq{DS1}, the parameters of the Lagrangian can be chosen in a way 
that the constant of our, A manifold, will have the correct sign. Effectively it was achieved  only by introduction in the Lagrangian
of the second sources for the scalar fields, the reasons and consequences of this construction must be clarified in the subsequent research. 
Of course, in the approach the constants are not really constants, we talk only about bare values of some infinite series for the
functions. Nevertheless, their finiteness in the flat space-time is related to the renormalizability of the corresponding effective action. Choosing a renormalizable 
field theory we will achieve a finite cosmological constant, that, of course does not guarantee it's smallness. 
We also notice, that in the framework the constants are result of the presence of classical solutions of the corresponding A and B fields equations of motion.
Namely, reformulating, the cosmological constant in this case this is an interaction of the condensate of scalar field 
(classical solution) with the manifold of interest trough the manifold's curvature. 

 Concerning the bare value of the constant, we speculate that the mostly natural value for the constant is zero at the case of flat space time.
This result can be provided immediately if we will consider the massless scalar fields, the bare cosmological constant will arise then as consequence of curvature corrections
to the flat propagators of the scalar fields. Another possibility for the zero bare values of the constants of A and B manifolds in the model with two sources is a precise
cancellation of the bare value due the cancellation of the sources in the expressions, in this case the mass of the scalar fields is arbitrary.
In general, in the case of the non-zero masses of the scalar fields, the constant's smallness
can be provided or by the super large mass of the fields or 
by very small values of the sources of the A and B field. In the first case, in order to provide the smallness
of the cosmological constants, the proposed mass must  be very heavy at least at the present epoch. 
It means, in general, that the propagation of the fields must be very small. In the second case, when the sources of the fields are very small, see \eq{Sml32}, 
the values of the sources are responsible for the smallness of the constant. 
Interesting issue, also, is a presence of the curvature in the final expression for the cosmological parameter through the effective action for the scalar fields. 
In this case, the effective action consists expressions with the powers of curvature higher then in ordinary Einstein equations. The construction
of these types of the terms can be performed on the base of gauge invariance of the action 
similarly to done in QCD and gravity at high energies, see \cite{EffA1}.

 In any case, the origin of the scalar field in the formalism is not clear. We can speculate that this field represents some
quantum fluctuations of the scalar curvature in the Einstein-Hilbert action which is not accounted by the classical equations of motion. 
The non-propagating of these kind of fields can be understood, for example, as result of their large masses.
This mass, in turn, leads to the definition of some new length which related to the plank mass and cosmological constant value combination. Of course, it is not clear, if 
this large mass mechanism is satisfactory in general and preferable in comparison with the framework with massless scalar fields. 
The cosmological constant, or parameter more precisely in the formalism, is evolving with the time and in order to understand the details of it's smallness some details of an evolution of the constant with the time must be clarified. In general, it is quite possible that the evolution begins from the small mass of the field or large curvature, or large constants correspondingly, and all parameters are decreasing with the time resulting in their now days values.

 In the article we also shortly discuss the possibilities of singularities crossing dictated by the formalism.
Our consideration of the subject is oversimplified, an extended application of the model to the problem we postpone for an another publication, nevertheless we notice the following. 
The symmetry of the model leads and dictates the special behavior of the mass parameter of the metric near the black hole singularity, in some extend the expansion of the parameter at vicinity of singularity is similar to the order parameter behavior in
Landau-Ginzburg theory. As consequence of this hypothetical expansion, which is nevertheless is dictated by the requests of the
analiticity of the metric and CPTM symmetry transform,  we obtain that the singularity disappears and instead we obtain some final value of the scalar curvature in $r\,=\,0$  point, this value is given by the $m_3$ expansion coefficient of the mass parameter.
Similarly, the Big Bang singularity at $t\,=\,0$ can be crossed analytically if we, following to the ideas of the paper, assume
that at initial moment of the time the only contribution to the Friedmann equation is the mutual interaction of the two parts
of the extended manifold. This interaction is anti-gravitational one, it is repulsion, it should decrease with the time ans as a consequence we obtain that there are analytical and exponentially growing scaling factors in FRW Universe metric. Of course, our 
calculations are far from complete. In general we need to consider the evolution of two manifolds related by CPTM symmetry, this mutual evolution can be different and depends on which type of the time paths in the manifolds we will choose. Also, the corresponding
scale factors will depend on the $\Lambda_{0}$ zero values for both manifolds, which interconnection can be quite complicated. Concerning the further time evolution of the $a(t)$ we have to account the consequent appear of the matter and radiation contributions in the equation as well as the proposition of the model that the cosmological constant is not constant but some complicated functional of interacting quantum fields which depend on time. Interesting to speculate and discuss, therefore, if that this "non-homogeneity"of the cosmological constant can be related to problem of Hubble tension, see for example \cite{Habble} and references therein.

 There are also the following issues which we did not discussed in the paper but which are arising naturally inside the framework. 
The first important question we can ask is about the renormalizability of the whole \eq{CC1} action.
Namely, an expansion of the full scalar propagators in the $S_{int}$ term of action will lead to the appearance
in the action new terms with different types of the dependence on the curvature tensor similarly to the higher curvature gravity theories.
This is an immediate consequence of the possible adiabatic expansion of the curved propagators directly in \eq{CC6} and, therefore,
the issue about a renormalizability of the \eq{CC1} action is an interesting one.

 Another issue is about the definition of the propagator of the scalar field of A manifold. In fact, for the hypothetical A and B scalar fields, we are not 
constrained by the choice of the Feynman propagator as the main one.
We can, for example, to choose the Wheeler propagator instead:
\beq\label{Concl1}
G_{W}\,=\,\frac{1}{2}\Le G_{F}\,-\,G_{D} \Ra\,,
\eeq
see \cite{Wheel,Wheel1}, it arises naturally in the \eq{DS5} expressions.
In the case of two types of scalar fields, this choice of A-field propagator will lead to the following propagator of the B-field:
\beq\label{Concl11}
CPTM(G_{W})\,=\,-\,\frac{1}{2}\Le G_{F}\,-\,G_{D} \Ra\,,\,\,\,\,\,CPTM(G_{W})\,=\,\frac{1}{2}\Le G_{F}\,-\,G_{D} \Ra\,
\eeq
for the first and second types of the field correspondingly. In the second case we obtain the totally symmetrical QFT,
a construction of the $in$-$in$ formalism for both cases is an interesting task.
An advantage of this choice, also, is that in this case the free quanta of the scalar fields
are absent and in some sense these particles are unobservable, see \cite{Wheel1}. It is an important property of the
field related to the cosmological constant and dark energy of course. The disadvantage of this choice is that 
there is no acceptable QFT based on this type of propagators, see discussion in \cite{Wheel1} anyway. Definitely, it is interesting problem for the further investigation.  

 The next question which we did not consider here is about the different definitions of bi-scalar functions in \eq{CC3}. There are a few additional possibilities 
exist. For example, instead the scalar fields we can consider any other fields with the similar 
overall zero contribution of the zero modes to the energy-stress tensor. In this set-up the problems with the origin of the fields and it's sources will arise as well of course 
and it will also request a clarification. 
Additionally, we can consider the bi-scalar functions in \eq{CC3} outside the QFT, there are plenty of possibilities exist for that. Nevertheless, in this case some different mechanisms of the smallness of the cosmological constant value must be established and clarified.

 Concluding we note that there are many different open questions that arise in the framework of the formalism and which investigation can help to understand 
and resolve the puzzle of the cosmological constant.

The author kindly acknowledges useful discussions of the subject with M.Zubkov.

\newpage
\section*{Appendix A: Propagators of scalar field }
\label{App1}
\renewcommand{\theequation}{A.\arabic{equation}}
\setcounter{equation}{0}

 For the simplicity we firstly calculate the Dyson propagator for the scalar filed $\phi$:
\beqar\label{A1}
G_{D}(x,y)& = &-\imath <\mathbf{\tilde{T}}\Le \phi(x)\,\phi(y)\Ra>=
-\imath \Le \theta(x^{0}-y^{0})<\phi(y)\,\phi(x)>\,+\,\theta(y^{0}-x^{0})<\phi(x)\,\phi(y)> \Ra=
\nonumber \\
&=&\,
\left\{ 
\begin{array}{c}
-\,\imath\,D(y\,-\,x)\,,\,\,\,x^{0}\,>\,y^{0}\,\\
-\,\imath\,D(x\,-\,y)\,,\,\,\,y^{0}\,>\,x^{0}\,
\end{array}
\right.\,.
\eeqar
Here
\beq\label{A9}
D(x\,-\,y)\,=\,<\phi(x)\,\phi(y)>\,=\,\int\,\frac{d^{3}k}{(2\pi)^{3}\,2\,\om_{k}}\,e^{-\imath\,\om_{k}\,(x^{0}-y^{0})+\imath\vec{k}(\vec{x}-\vec{y})}
\eeq
is Wightman function.
Using the $\theta$ function representation
\beq\label{A2}
\theta(x^{0}-y^{0})\,=\,\frac{\imath}{2\pi}\,\int\,d \om\,\frac{e^{-\imath\,\om\,(x^{0}-y^{0})}}{\om\,+\,\imath\,\varepsilon}
\eeq
we obtain for the first term in the r.h.s. of \eq{A1}:
\beq\label{A3}
\int\,d\om\,\frac{e^{-\imath\,\om\,(x^{0}-y^{0})}\,e^{\imath\,\om_{k}\,(x^{0}-y^{0})}}{\om\,+\,\imath\,\varepsilon}\,
\int\,\frac{d^{3}k}{(2\pi)^{4}\,2\,\om_{k}}\,e^{-\imath\vec{k}(\vec{x}-\vec{y})}
\eeq
with 
\beq\label{A4}
<\phi^{-}(k)\,\phi^{+}(k^{'})>\,=\,\delta_{k\,k^{'}}\,,\,\,\,\om_{k}\,=\,\sqrt{k^{2}\,+\,m^2}\,.
\eeq
After the variables change 
\beq\label{A5}
\om\,-\,\om_{k}\,\rightarrow\,\om\,,\,\,\,\vec{k}\,\rightarrow\,-\vec{k}
\eeq
we obtain for this term:
\beq\label{A6}
\int\,\frac{d^{4}k}{(2\pi)^{4}\,2\,\om_{k}}\,\frac{e^{-\imath\,\om\,(x^{0}-y^{0})+\imath\vec{k}(\vec{x}-\vec{y})}}{\om\,+\,\om_{k}\,+\,\imath\,\varepsilon}\,=\,
\int\,\frac{d^{4}k}{(2\pi)^{4}\,2\,\om_{k}}\,\frac{e^{-\imath\,k\,(x-y})}{\om\,+\,\om_{k}\,+\,\imath\,\varepsilon}\,.
\eeq
For the second one we have:
\beq\label{A7}
-\,\int\,\frac{d^{4}k}{(2\pi)^{4}\,2\,\om_{k}}\,\frac{e^{-\imath\,k\,(x-y})}{\om\,-\,\om_{k}\,-\,\imath\,\varepsilon}\,
\eeq
that provides for the propagator:
\beq\label{A8}
G_{D}(x,y)\,=\,-\,\int\,\frac{d^{4}k}{(2\pi)^{4}\,}\,\frac{e^{-\imath\,k\,(x-y})}{k^2\,-\,m^2\,-\,\imath\,\varepsilon}\,.
\eeq
The similar calculations provide for the Feynman propagator
\beqar\label{A10}
G_{F}(x,y)\,& = &-\imath <\mathbf{T}\Le \phi(x)\,\phi(y)\Ra>=
-\imath\,\Le \theta(x^{0}-y^{0})<\phi(x)\,\phi(y)>+\theta(y^{0}-x^{0})<\phi(y)\,\phi(x)> \Ra=\nonumber \\
&=&
\left\{ 
\begin{array}{c}
-\,\imath\,D(x\,-\,y)\,,\,\,\,x^{0}\,>\,y^{0}\,\\
-\,\imath\,D(y\,-\,x)\,,\,\,\,y^{0}\,>\,x^{0}\,
\end{array}
\right.\,
\eeqar
the following answer:
\beq\label{A11}
G_{F}(x,y)\,=\,\int\,\frac{d^{4}k}{(2\pi)^{4}\,}\,\frac{e^{-\imath\,k\,(x-y})}{k^2\,-\,m^2\,+\,\imath\,\varepsilon}\,.
\eeq

\newpage
\section*{Appendix B: Propagator of scalar field in curved manifold}
\renewcommand{\theequation}{B.\arabic{equation}}
\setcounter{equation}{0}

 We begin from the usual definition of the quadratic with respect to the fluctuations part of the \eq{CC4} Lagrangian:
\beq\label{B1}
L_{\varepsilon^2}\,=\,\sqrt{-g}\,\Le\,\varepsilon\,G_{F}^{-1}\,\varepsilon\,\Ra\,=\,
-\,\sqrt{-g}\,\Le\,\varepsilon\,\Box\,\varepsilon\,\Ra
\eeq
with
\beq\label{B2}
\Box\,=\,\frac{1}{\sqrt{-g}}\,\D_{\nu}\,\Le\,\sqrt{-g}\,g^{\nu \mu}\D_{\mu}\,\Ra\,+\,m^2\,=\,\D_{\mu}\D^{\,\mu}\,+\,m^2\,+\,M_{1}
\eeq
and following formal definition of the Green's function:
\beq\label{B3}
\Box_{x}\,G(x,y)\,=\,\frac{1}{\sqrt{-g(x)}}\,\Box_{x}\,\Le\,\D_{\mu}\D^{\,\mu}\,+\,m^2\,+\,M_{1}\,\Ra^{-1}_{x y}\,=\,-\,
\frac{1}{\sqrt{-g(x)}}\,\delta^{4}(x-y)\,
\eeq
which we can rewrite as
\beqar\label{B4}
\sqrt{-g(x)}\,\Box_{x} G(x,y) & = &\Le\, \D_{\nu}\,\Le\,\sqrt{-g}\,g^{\nu \mu}\D_{\mu}\,\Ra\,+\,m^2\,\sqrt{-g}\,\Ra_{x} G(x,y)= 
\nonumber \\
&=&\Le\D_{\mu}\D^{\,\mu}\,+\,m^2\,+\,N_{1}\Ra_{x} G(x,y)=-\delta^{4}(x-y)\,.
\eeqar
From \eq{B2} and \eq{B4} we have the following definition of $N_{1}$ operator:
\beq\label{B5}
N_1 = \Le \sqrt{-g} g^{\mu \nu} - \eta^{\mu \nu}\Ra \D_{\mu}  \D_{\nu} + \Le \sqrt{-g} - 1\Ra m^{2} +
\sqrt{-g} \Le \D_{\mu} g^{\mu \nu} \Ra \D_{\nu} - \frac{\sqrt{-g}}{2} g^{\mu \nu} g_{\rho \sigma} \Le \D_{\mu} g^{\rho \sigma} \Ra \D_{\nu} \,.
\eeq
The precise perturbative solution of the propagator of scalar field in the curved space time is given, thererefore, by the following expression:
\beq\label{B6}
G(x,y)\,=\,G_{0}(x,y)\, + \,\int\, d^4 z\,G_{0}(x,z)\,N_{1}(z)\,G(z,y)\,
\eeq
with
\beq\label{B7}
\Le \D_{\mu}\D^{\,\mu}\,+\,m^2\Ra_{x}\,G_{0}(x,y)\,=\,-\,\delta^{4}(x-y)\,.
\eeq
Here we do not define precisely which $G_{0}(x,y)$ propagator to use in \eq{B6}, it must satisfy \eq{B7} with arbitrary boundary conditions.
The \eq{B6} representation of the propagator is usefull due the few reasons. First of all, it can be used in the weak field approximation, expanding 
$N_1$ operator in terms of $h^{\mu \nu}$ we will obtain perturbative expression for the propagator with required precision. Another interesting application
of \eq{B6} is to use it as a reformulatedF recursive formula for the adiabatic expansion of the $G(x,y)$, see \cite{Park}. In this case, inserting
the adiabatic series in both sides of \eq{B6}, we will obtain some non-local relations between the $a_{l}(x,y)$ coefficients of the adiabatic expansion.

 In the formalism we can use the \eq{B6} for the calculation of the variation of the cosmological constant with respect to the metric. 
The non-trivial variation of the interaction part of the action, \eq{CC6}, is provided by the variation of the full propagator in the curved space time.
We have for this variation:
\beq\label{B8}
\delta\,G(x,y)\,=\,\int\, d^4 z\,G_{0}(x,z)\,\Le \delta\,N_{1}(z)\Ra \,G(z,y)\,+\,\int\, d^4 z\,G_{0}(x,z)\,N_{1}(z)\,\Le \delta\, G(z,y)\Ra \,.
\eeq
The same we can rewrite as:
\beq\label{B9}
\int\, d^4 z\,\Le \delta^4 (x-z)\,-\, G_{0}(x,z)\,N_{1}(z)\Ra\,\delta\, G(z,y)\,=\,\int\, d^4 z\,G_{0}(x,z)\,\Le \delta\,N_{1}(z)\Ra \,G(z,y)\,
\eeq
or
\beqar\label{B91}
&\,&
\int\, d^4 x\,\Le \delta^4 (p-x)\,-\, G_{0}(p,x)\,N_{1}(x)\Ra^{-1}\,
\int\, d^4 z\,\Le \delta^4 (x-z)\,-\, G_{0}(x,z)\,N_{1}(z)\Ra\,\delta\, G(z,y)\,=\,
\nonumber \\
&=&\,
\int\, d^4 x\,\Le \delta^4 (p-x)\,-\, G_{0}(p,x)\,N_{1}(x)\Ra^{-1}\,
\int\, d^4 z\,G_{0}(x,z)\,\Le \delta\,N_{1}(z)\Ra \,G(z,y)\,
\eeqar
that provides finally:
\beq\label{B10}
\delta\, G(x,y)\,=\,
\int\, d^4 p\,\int\, d^4 z\,\Le \delta^4 (x-p)\,-\, G_{0}(x,p)\,N_{1}(p)\Ra^{-1}\,
G_{0}(p,z)\,\Le \delta\,N_{1}(z)\Ra \,G(z,y)\,.
\eeq
Taking variation of the $N_{1}$ and assuming that the metric does not depend on the cosmological constant, we obtain:
\beqar\label{B11}
&\,& \delta\, G(x,y)\,= \,\int\, d^4 p\,\int\, d^4 z\,\Le \delta^4 (x-p)\,-\, G_{0}(x,p)\,N_{1}(p)\Ra^{-1}\,G_{0}(p,z)\,\delta\,g^{\mu \nu}(z)\,
\sqrt{-g(z)}\,
\nonumber \\
&\,&
\Le\, - \frac{1}{2}  g_{\mu \nu} 
\Le g^{w p} \D_{w}\D_{p} + m^2 +(\D_{\rho} g^{\rho \sigma})\D_{\sigma} - \frac{1}{2} g^{w p} g_{\rho \sigma} (\D_{w} g^{\rho \sigma}) \D_{p} \Ra +
\right.\nonumber \\
&+&
\left.
\Le \D_{\mu}\D_{\nu} - \frac{1}{2} g_{\rho \sigma} (\D_{\mu} g^{\rho \sigma}) \D_{\nu} 
+ \frac{1}{2} g^{\rho \sigma} (\D_{\rho} g_{\mu \nu}) \D_{\sigma} \Ra \,\Ra\,G(z,y)\,+\,
\nonumber \\
&+&
\int\, d^4 p\,\int\, d^4 z\,\Le \delta^4 (x-p)\,-\, G_{0}(x,p)\,N_{1}(p)\Ra^{-1}\,\delta g^{\mu \nu}(z) \,
\nonumber \\
&\,&
\Le\,-\, \D_{\mu,z}
\Le \sqrt{-g}\,G_{0}(p,z)\,\D_{\nu, z} G(z,y)\Ra +
\frac{1}{2} \, \D_{\rho,z}
\Le \sqrt{-g}\,g^{\rho \sigma}\, g_{\mu \nu}\, G_{0}(p,z)\,\D_{\sigma, z} G(z,y)\Ra \Ra
\eeqar
or
\beqar\label{B111}
&\,& \delta\, G(x,y)\,= \,\frac{1}{2}\int\, d^4 p\,\Le \delta^4 (x-p)\,-\, G_{0}(x,p)\,N_{1}(p)\Ra^{-1}\,G_{0}(p,y)\,g_{\mu \nu}(y) \,\delta\,g^{\mu \nu}(y)\,
+
\nonumber \\
&+&
\int\, d^4 p\,\int\, d^4 z\,\Le \delta^4 (x-p)\,-\, G_{0}(x,p)\,N_{1}(p)\Ra^{-1}\,G_{0}(p,z)\,\delta\,g^{\mu \nu}(z)\,
\sqrt{-g(z)}\,
\nonumber \\
&\,&
\Le \D_{\mu}\D_{\nu} - \frac{1}{2} g_{\rho \sigma} (\D_{\mu} g^{\rho \sigma}) \D_{\nu} 
+ \frac{1}{2} g^{\rho \sigma} (\D_{\rho} g_{\mu \nu}) \D_{\sigma} \Ra \,G(z,y)\,+\,
\nonumber \\
&+&
\int\, d^4 p\,\int\, d^4 z\,\Le \delta^4 (x-p)\,-\, G_{0}(x,p)\,N_{1}(p)\Ra^{-1}\,\delta g^{\mu \nu}(z) \,
\nonumber \\
&\,&
\Le\,-\, \D_{\mu,z}
\Le \sqrt{-g}\,G_{0}(p,z)\,\D_{\nu, z} G(z,y)\Ra +
\frac{1}{2} \, \D_{\rho,z}
\Le \sqrt{-g}\,g^{\rho \sigma}\, g_{\mu \nu}\, G_{0}(p,z)\,\D_{\sigma, z} G(z,y)\Ra \Ra\,.
\eeqar
Now we can write the new term in the equations of motion for the gravitational field. Taking variation 
of the cosmological constant term in $S_{int}$ for the A minifold, \eq{CC6}, with $f=1$ source we obtain:
\beqar\label{B12}
\delta\,S_{int} & = &
\frac{m_{p}^{6}}{2}\int d^4 x\int d^4 y\,\sqrt{-g(x)}\Le  g_{\mu \nu} \delta g^{\mu \nu}\Ra G(x,y)\,\sqrt{-g(y)}\,-
\nonumber \\
&-&
\frac{m_{p}^{6}}{2}\int d^4 x\int\, d^4 y\,\sqrt{-g(x)}\Le \delta\,G(x,y)\Ra\sqrt{-g(y)}\,+\,\cdots
\eeqar
with similar contribution for B manifold added. If we introduce in $N_1$ expression a dependence of the metric on the cosmological constant,
then we will need to take into account in the variation also the following terms
\beq\label{B13}
\frac{\delta }{\delta g^{\mu \nu}}\,\Le A_{1}(g)\,\Lambda\,+\,A_{2}(g)\,\Lambda^{2}\,+\,\cdots \Ra\delta g^{\mu \nu} 
\eeq
which arise in the variation after an expansion of the operator with expect to the constant in the form of perturbative series.

\newpage
\section*{Appendix C: Integrals of scalar propagators}
\label{App3}
\renewcommand{\theequation}{C.\arabic{equation}}
\setcounter{equation}{0}

 There are the following integrals we need to calculate:
\beq\label{C1}
I_{F}\,=\,\int\, d^4 y \,G_{F}(x,y)\,,\,\,\,
I_{D}\,=\,\int\, d^4 y \,G_{D}(x,y)\,.
\eeq
For the first integral, using an equivalent to \eq{A11} representation of the bare Feynman propagator, we can write:
\beqar\label{C2}
I_{F}& = &-\imath\,\int d^4 y\,\theta(x^0 - y^0)\,\int\,\frac{d^{3} k}{2\om_{k}(2\pi)^{3}}\,e^{-\imath\,(\om_{k}\,-\,\imath\varepsilon)\,(x^0 - y^0)\, + 
\imath\,\vec{k}\,(\vec{x}-\vec{y})} -
\nonumber \\
&-&\,
\imath\,\int d^4 y\,\theta(y^0 - x^0)\,\int\,\frac{d^{3} k}{2\om_{k}(2\pi)^{3}}\,e^{-\imath\,(\om_{k}\,-\,\imath\varepsilon)\,(y^0 - x^0) + 
\imath\,\vec{k}\,(\vec{y}-\vec{x})}\,=\,
\nonumber \\
&=&
\imath\,\int_{\infty}^{-\infty} d y^0\,\theta(y^0 )\,\int\,d^3 y\,e^{\imath\,\vec{k}\,\vec{y}}\,
 \int\,\frac{d^{3} k}{2\om_{k}(2\pi)^{3}}\,e^{-\imath\, (\om_{k}\,-\,\imath\varepsilon)\,y^0 \,}\,-\,
\nonumber \\
&-&\imath\,\int_{-\infty}^{\infty} d y^0\,\theta(y^0 )\,\int\,d^3 y\,e^{\imath\,\vec{k}\,\vec{y}}\,
 \int\,\frac{d^{3} k}{2\om_{k}(2\pi)^{3}}\,e^{-\imath\, (\om_{k}\,-\,\imath\varepsilon)\,y^0 \,}\,
 =
\nonumber \\
&=&\,
-2\,\imath\,\int dy^0\,\theta(y^0 )\,\int\,\frac{d^{3} k}{2\om_{k}}\,e^{-\imath\, (\om_{k}\,-\,\imath\varepsilon)\,y^0}\,\delta^{3}(k)\,=\,
-\frac{\imath}{|m|}\,\int_{0}^{\infty}\,d y^{0}\,e^{-\imath\, (|m|\,-\,\imath\varepsilon)\,y^0}\,=\,
\nonumber \\
&=&\,
-\frac{1}{|m|\,(|m|\,-\,\imath\varepsilon)}\,.
\eeqar
The same can be done for $I_{D}$, we will obtain then:
\beq\label{C3}
I_{D}\,=\,-\,I^{*}_{F}\,=\,\frac{1}{|m|\,(|m|\,+\,\imath\varepsilon)}\,.
\eeq

\end{document}